\documentclass{aa}  

\usepackage{graphicx}
\usepackage{txfonts}
\usepackage{hyperref}
\hypersetup{
    colorlinks=true,
    linkcolor=cyan,
    citecolor=cyan,
    urlcolor=cyan,
    }

\begin{document}

   \title{Revisiting symbiotic binaries with interferometry}

    \subtitle{I. The PIONIER archival collection}
    
   \author{Jaroslav Merc
          \inst{1,2,3}
          \and
          Henri M.J. Boffin\inst{1}
          }

   \institute{European Southern Observatory, Karl-Schwarzschild-Stra\ss{}e 2, Garching bei M\"{u}nchen, 85748, Germany
         \and
             Astronomical Institute of Charles University, V Hole\v{s}ovi\v{c}k{\'a}ch 2, Prague, 18000, Czech Republic
             \and
             Instituto de Astrof\'isica de Canarias, Calle Vía Láctea, s/n, E-38205 La Laguna, Tenerife, Spain
             }

   \date{Received January 17, 2025; accepted January 28, 2025}

 
  \abstract
   {Symbiotic stars serve as exceptional laboratories for investigating mass transfer processes in binary systems. However, the dominant mechanism of mass transfer from the red giant donor to the compact accretor  —  typically a white dwarf or, in rare cases, a neutron star  —  remains unclear. It is uncertain whether it is driven primarily by the stellar wind, Roche-lobe overflow, or a combination of the two. While radii inferred from rotational velocities or spectral types suggest smaller Roche-lobe filling factors, the presence of ellipsoidal variability, presumably caused by tidally deformed giants in many symbiotic systems, indicates the opposite. Interferometric observations of symbiotic giants, combined with distance measurements provided by the \textit{Gaia} mission, offer a promising avenue to resolve this discrepancy. In this first paper of the series, we (re)analyze VLTI/PIONIER observations of six symbiotic stars: AG Peg, FG Ser, ER Del, V1261 Ori, RW Hya, and V399 Pav. With the exception of the uncertain case of V399 Pav, we find that the giants in these systems remain well within their canonical Roche lobes, even in V1261 Ori and RW Hya, where ellipsoidal variability is observed. All six stars appear to be rather luminous and likely located on the asymptotic giant branch, although the possibility of some of them being at the tip of the first red giant branch cannot be ruled out.}

   \keywords{binaries: symbiotic -- stars: mass-loss -- techniques: interferometric
               }

   \maketitle
%

\section{Introduction}

Symbiotic stars are ideal for studying mass transfer mechanisms as they are among the widest interacting binaries, with orbital periods ranging from hundreds to thousands of days \citep[e.g.,][]{2013AcA....63..405G,2019RNAAS...3...28M}. These systems consist of a cool giant (typically of spectral type M, or less frequently, K) and a hot compact star, usually a white dwarf or, in some cases, a neutron star \citep[see, e.g.,][]{2012BaltA..21....5M,2019arXiv190901389M}. One critical unresolved aspect pertains to the mechanism responsible for mass transfer from the giant to its companion -- whether it is dominated by Roche-lobe overflow (RLOF), stellar wind, or a combination of the two\footnote{It is clear that the symbiotic giants in so-called D-type symbiotic stars (dusty infrared types) that host very evolved Mira variables with orbital periods on the order of tens of years cannot fill their Roche lobes, but this cannot be stated a priori for the S-type (stellar infrared type) symbiotic stars we are interested in.}.

Initially, it seemed that most symbiotic giants did not fill their Roche lobes, as their radii -- estimated from spectral types or rotational velocities -- only corresponded to approximately 0.4–0.5 of their respective Roche lobes, and the ellipsoidal effect, attributed to tidally distorted giants, had previously been observed in only a small sample of symbiotic stars, all indicative of prevalent wind-driven mass transfer. However, subsequent analysis of infrared light curves of several S-type symbiotic binaries with orbital periods of less than $\approx$ 1000 days revealed a substantially higher occurrence of ellipsoidal light curve variations \citep[e.g.,][]{2007BaltA..16...49R,2012BaltA..21....5M}. 

Significant challenges arise for certain symbiotic stars that exhibit ellipsoidal variability, indicating that the giant might be filling its Roche lobe. In some cases, the inferred radii derived from rotational velocity are notably smaller than the expected size of their Roche lobes \citep[the ``rotation-$v_{\rm rot} \sin{i}$ problem'';][]{2012BaltA..21....5M}. Given the relatively short timescales expected for synchronization, it seemed improbable that this discrepancy results from asynchronous rotation. Another possibility is that the dense, slow wind is filling the Roche lobe, rather than the star itself, or that the modification of the Roche potential of a pulsating star with strong wind might explain the discrepancy. An additional problem in these systems is that, given the mass ratio, many such objects should undergo unstable mass transfer according to current evolutionary theories, and it is unlikely that we are fortunate enough to observe all of them in the brief episode at the beginning of this phase.

The most straightforward test is, clearly, a direct measurement of the radii of symbiotic giants through interferometric techniques. 
By leveraging interferometrically measured angular diameters in conjunction with distances now available thanks to the \textit{Gaia} mission \citep{2016A&A...595A...1G}, one can obtain accurate linear radii and, consequently, Roche-lobe filling fractions for a substantial sample of bright and nearby symbiotic giants. Here, we searched the ESO archive\footnote{archive.eso.org} and the Jean-Marie Mariotti Center Optical Interferometry DataBase\footnote{http://oidb.jmmc.fr} for interferometric observations of all confirmed symbiotic stars from the New Online Database of Symbiotic Variables \citep{2019RNAAS...3...28M,2019AN....340..598M} obtained using the Precision Integrated-Optics Near-infrared Imaging ExpeRiment (PIONIER) instrument on ESO's Very Large Telescope Interferometer (VLTI). Observations for six known symbiotic systems were retrieved -- AG~Peg, FG~Ser, ER Del, V1261 Ori, RW Hya, and V399 Pav. These systems were mostly observed during one epoch, except for V399 Pav, which had two epochs obtained approximately 6 weeks apart, and FG~Ser, which had multiple visits. 

Data for two stars from this list, RW Hya and V399 Pav, have never been analyzed before. Observations of the other four\footnote{Additional objects analyzed in \citet{2014A&A...564A...1B}, namely HD 352 and V1472 Aql, or SS Lep \citep{2011A&A...536A..55B} do not fulfill the traditional symbiotic criteria and are therefore not included here.} were previously discussed in \citet{2014A&A...564A...1B}, but we extend this analysis with new interferometric data and/or additional assumptions, such as considering more realistic limb-darkened models instead of a simple uniform disk. Furthermore, the accuracy of the results obtained by \citet{2014A&A...564A...1B}, particularly the linear radii, was significantly limited by the imprecise or unknown distances to these objects at that time. The data published in the recent \textit{Gaia} Data Release 3 \citep[DR3;][]{2023A&A...674A...1G} now offer a unique opportunity to, at least partly, overcome this limitation.

We find that all of the symbiotic stars we studied appear to be filling only part of their Roche lobe and that they are most likely all on the asymptotic giant branch (AGB; in two cases, likely even in the thermally pulsing phase). The derived Roche-lobe filling factors imply that mass transfer likely occurs via a stellar wind or wind-RLOF. However, the cause of the observed ellipsoidal variability in two of these objects remains unclear under these conditions. Alternatively, the Roche lobe size may need to be less than its canonical value to permit RLOF. Yet, current mass transfer theories indicate that such a scenario would result in unstable mass transfer, significantly shortening the duration of this evolutionary stage.

\section{Observations, data reduction, and analysis}
In this work we analyzed archival interferometric data of six confirmed symbiotic stars, obtained using the PIONIER instrument \citep{2011A&A...535A..67L} on the 1.8 m Auxiliary Telescopes (ATs) of the ESO's Very Large Telescope. Most observations utilized all four ATs, with the exception of one epoch for FG~Ser, where only three baselines were available. The data were collected between March 2012 and May 2019. A detailed log of individual observations is provided in Table \ref{tab:log}.  

\begin{table*}[h]%
\centering
\caption{VLTI/PIONIER observations of the target stars and the corresponding angular diameters obtained in this work.\label{tab:log}}%
\begin{tabular}{lcccclcccc}
\hline
\noalign{\smallskip}
Object & $J$ & $H$ & $K_s$ & Night & Configuration & UD & $\chi^2$ & LDD & $\chi^2$\\
 & [mag] & [mag] & [mag] &  &  & [mas] &  & [mas] & \\
\noalign{\smallskip}
\hline
\noalign{\smallskip}
AG~Peg & 5.0 & 4.4 & 3.9 & 2012-08-13 & A1-G1-I1-K0  & 1.023$\pm$0.009 & 0.745 & 1.062$\pm$0.009 & 0.744\\
FG~Ser & 5.9 & 4.9 & 4.4 & 2012-07-03 & A1-G1-I1-K0  &  0.813$\pm$0.009 & 1.358 & 0.845$\pm$0.009 & 1.358 \\
 & & & & 2012-08-13 & A1-G1-I1-K0  &  0.941$\pm$0.015 & 0.206 & 0.978$\pm$0.016 & 0.206  \\
 & & & & 2012-08-20 & A1-I1-K0  &  0.892$\pm$0.011 & 0.548 & 0.927$\pm$0.012 & 0.549  \\
 & & & & 2014-08-27 & A1-G1-J3-K0   &  0.853$\pm$0.004 & 0.571 & 0.887$\pm$0.005 & 0.571  \\
 & & & & 2014-09-22 & A1-G1-J3-K0  &  0.857$\pm$0.002 & 0.439 & 0.891$\pm$0.002 & 0.438   \\
 & & & & 2019-04-29 & A0-D0-G1-J3  &  0.887$\pm$0.008 & 0.745 & 0.922$\pm$0.009 & 0.744   \\
 & & & & 2019-05-02 & A0-G1-J2-J3  &  0.885$\pm$0.004 & 0.252 & 0.920$\pm$0.004 & 0.252   \\
 & & & & 2019-05-03 & A0-G1-J2-J3  &  0.894$\pm$0.003 & 0.310 & 0.929$\pm$0.003 & 0.309   \\
ER Del & 6.2 & 5.3 & 5.0 & 2012-08-13 & A1-G1-I1-K0  & 0.613$\pm$0.009 & 0.150 & 0.636$\pm$0.009 & 0.150  \\
V1261 Ori & 3.3 & 2.4 & 2.1 & 2012-03-02 & A1-G1-I1-K0  &  2.292$\pm$0.003 & 0.531 & 2.383$\pm$0.003 & 0.533 \\
RW Hya & 5.7 & 4.9 & 4.6 & 2014-08-27 & A1-G1-J3-K0  &   0.619$\pm$0.010 & 0.317 & 0.642$\pm$0.010 & 0.317\\
V399 Pav & 6.7 & 5.8 & 5.4 & 2014-08-27 & A1-G1-J3-K0  &   0.517$\pm$0.016 & 0.501 & 0.536$\pm$0.017 & 0.501   \\
 & & & & 2014-10-06 & A1-G1-J3-K0  &   0.493$\pm$0.031 & 0.581 & 0.511$\pm$0.033 & 0.581    \\
\hline
\end{tabular}
\tablefoot{Angular diameters are provided for both the uniform disk (UD) and limb-darkened disk (LDD) models. The infrared magnitudes for the analyzed stars shown in the Table are from 2MASS \citep{2006AJ....131.1163S}.}
\end{table*}

 \begin{figure}
   \centering
   \includegraphics[width=0.7\columnwidth]{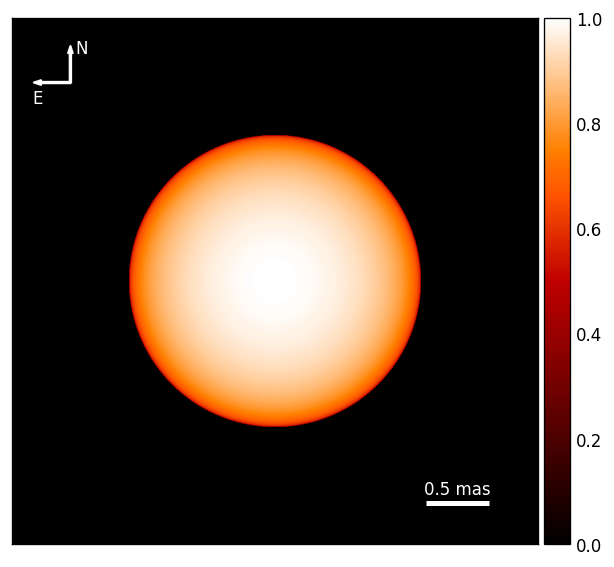}
      \caption{Illustration of the model used for fitting the interferometric observations. The particular limb-darkening coefficients and angular diameter are those for V1261 Ori.
              }
         \label{fig:model}
   \end{figure}

Data reduction was carried out using the standard {\tt pndrs} package \citep{2011A&A...535A..67L}, which yielded squared visibilities and closure phases for all epochs. For all targets, the closure phases were consistent with zero within their respective errors, indicating no significant asymmetries in the observed stars. Consequently, only the squared visibilities were used in the subsequent analysis.

To derive the angular diameters of the target stars, we employed the PMOIRED code \citep{2022SPIE12183E..1NM} and tested two models: the uniform disk model and the limb-darkened model (see an example in Fig.~\ref{fig:model}). In both models, the sole fitted parameter was the angular diameter. However, the limb-darkened model is considered to provide a more realistic flux distribution and, therefore, more accurate results. The difference between the models was minimal (the average correction factor for our sample was approximately 1.04), and the improvement in the reduced $\chi^2$ was negligible. We adopted the power-2 limb-darkening law of \citet{2023A&A...674A..63C}, with coefficients corresponding to the T$_{\rm eff}$ and $\log g$ values of individual stars, as listed in Table \ref{tab:target_stars}.

\begin{figure*}
   \centering
   \includegraphics[width=0.66\columnwidth]{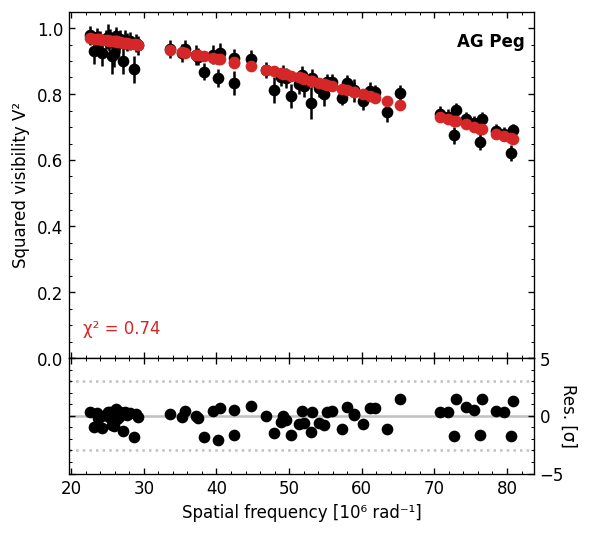}
   \includegraphics[width=0.66\columnwidth]{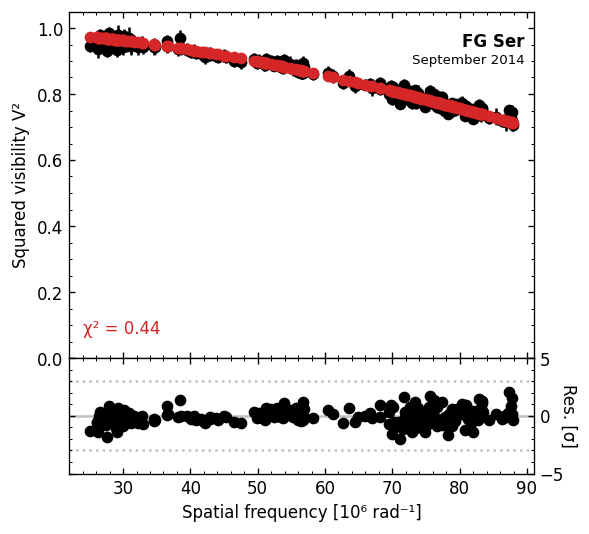}
   \includegraphics[width=0.66\columnwidth]{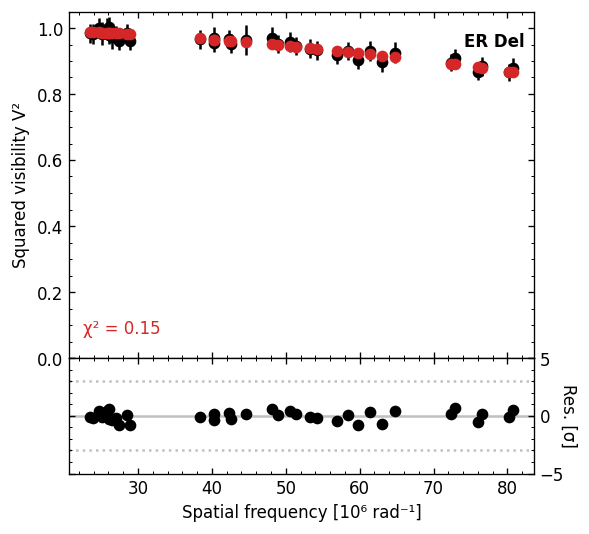}
   \includegraphics[width=0.66\columnwidth]{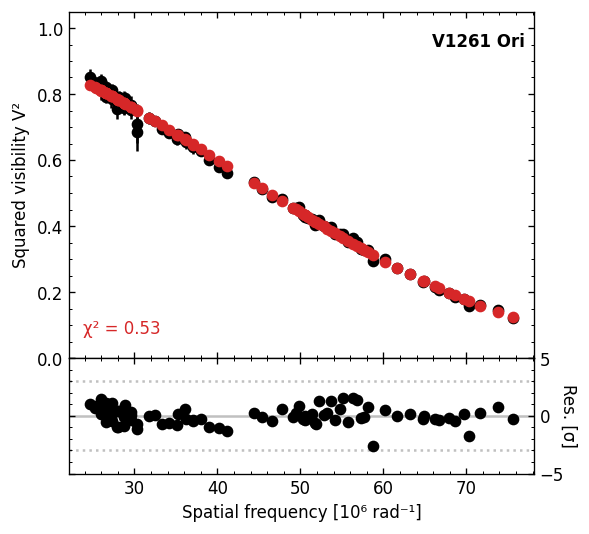}
   \includegraphics[width=0.66\columnwidth]{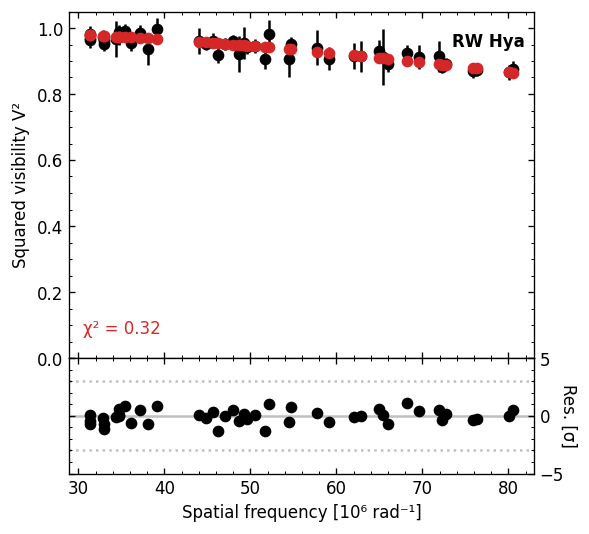}
   \includegraphics[width=0.66\columnwidth]{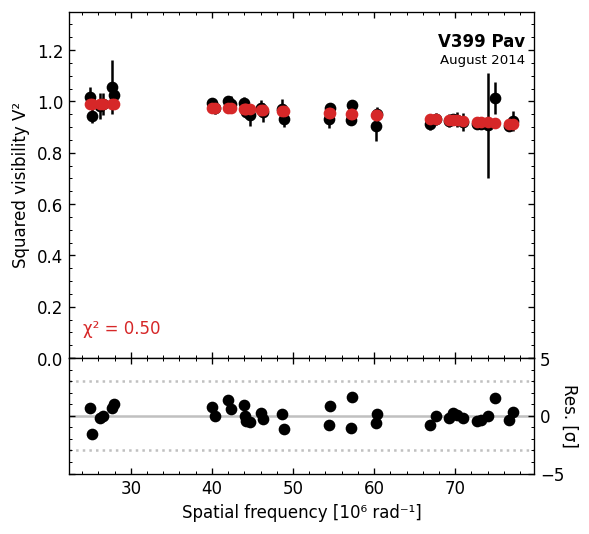}
      \caption{VLTI/PIONIER squared visibilities as a function of spatial frequency for the target stars. Observed data are represented in black, with the best-fitting model displayed in red. For stars with multiple observations, only one epoch is shown.
              }
         \label{fig:fit}
   \end{figure*}

The resulting angular diameters are provided in Table \ref{tab:log}, with the observed data and corresponding best-fit models shown in Fig.~\ref{fig:fit}. The reduced $\chi^2$ values confirm that the data were well-fitted using symmetric models\footnote{We note that in some cases, the $\chi^2$ values are relatively low, suggesting that the PIONIER pipeline may overestimate the uncertainties in the interferometric observables. As a result, the reported uncertainties in radii might also be overestimated. However, this has little impact on the results, as the primary source of uncertainty of the radii arises from the distance estimates.}. We note that our analysis assumes the contribution of the companion in the $H$ band is negligible (a reasonable assumption for symbiotic stars, as their hot components primarily emit in the UV and X-ray wavelengths), and that there is no background flux in this wavelength range (e.g., from circumstellar material), as suggested by the analysis of \citet{2014A&A...564A...1B}.

\begin{table*}%
\centering
\caption{Effective temperatures and surface gravities of the cool components of the target stars used in the limb-darkening model, along with their orbital elements.\label{tab:target_stars}}%
\begin{tabular}{lllcllllc}
\hline
\noalign{\smallskip}
Object & T$_{\rm eff}$ & $\log g$ & Ref. & $P$ & $e$ & $f$(m) & Ref.\\
 & [K] & [dex] &   & [days] &  & [M$_\odot$]&\\
 \noalign{\smallskip}
\hline
\noalign{\smallskip}
AG~Peg & 3500 & 0.5 & 1 & 818.2$\pm$1.6 & 0.11$\pm$0.04 & 0.0135$\pm$0.0015 & 2\\
FG~Ser & 3400 & 0.5 & 1 & 633.5$\pm$2.4 & 0 & 0.0218$\pm$0.0025 & 3\\
ER Del$^a$ & 3500 & 0.5 & 4 &2081$\pm$2 & 0.281$\pm$0.003  & 0.0689$\pm$0.002 & 5 \\
V1261 Ori$^a$ & 3500  & 0.5 & 4 &638.1$\pm$0.1 & 0.061$\pm$0.003  & 0.0454 & this work  \\
RW Hya & 3700 & 0.5 &  6 &370.4$\pm$0.8 & 0 & 0.026$\pm$0.004 & 7 \\
V399 Pav & 3300 & 0.0 &  6 &560 & 0.05$\pm$0.04 & 0.022 & 8 \\
\hline
\end{tabular}
\tablefoot{$^a$The giant in this system is an S star.}
\tablebib{(1)~\citet{2023MNRAS.526..918G}; (2) \citet{2000AJ....119.1375F}; (3) \citet{2000AJ....120.3255F}; (4) \citet{2014A&A...564A...1B}; (5) \citet{2019A&A...626A.127J}; (6) \citet{2016MNRAS.455.1282G}; (7) \citet{1996A&A...306..477S}; (8) \citet{2006BAAA...49..132B}.
}
\end{table*}

\section{Results and discussion}

\begin{table*}%
\centering
\caption{\textit{Gaia} DR3 \citep{2023A&A...674A...1G} data for the target stars.}%
\begin{tabular}{lllllllllcll}
\hline
\noalign{\smallskip}
Object & $BP$ & $G$ & $RP$ & GOF$^a$ & RUWE$^b$ & $\varpi$ & Dist. (inv.) & Dist. (B-J) & Dist. (lit.) & Ref. & \\
 & [mag] & [mag] & [mag] &  &  & [mas] & [pc] & [pc] & [pc] &  \\
 \noalign{\smallskip}
\hline
\noalign{\smallskip}
AG~Peg & 8.8 & 7.8 & 6.7 & 9.56 & 1.36 & 0.754$\pm$0.035 & 1326$^{+65}_{-59}$ & 1272$^{+51}_{-49}$ & 800 & 1\\
\noalign{\smallskip}
FG~Ser & 11.9 & 9.7 & 8.4 & 1.06 & 1.05 & 0.625$\pm$0.046 & 1600$^{+127}_{-110}$ & 1472$^{+112}_{-90}$ & 1100$\pm$200 & 2 \\
\noalign{\smallskip}
ER Del & 10.5 & 9.0 & 7.9 & 1.24 & 1.04 & 0.483$\pm$0.025 & 2070$^{+113}_{-102}$ & 1903$^{104}_{-89}$ & 1750 - 2600 & 3\\
\noalign{\smallskip}
V1261 Ori & 7.2 & 5.8 & 4.7 & 69.75 & 4.77 & 2.617$\pm$0.125 & 382$^{+19}_{-17}$ & 375$^{+17}_{-17}$ & 510$^{+375}_{-215}$ & 3 \\
\noalign{\smallskip}
RW Hya & 9.0 & 8.1 & 7.1 & 6.79 & 1.27 & 0.648$\pm$0.035 & 1543$^{+88}_{-79}$ & 1479$^{+67}_{-81}$ & 820$\pm$112 & 4 \\
\noalign{\smallskip}
V399 Pav & 11.5 & 9.9 & 8.7 & -1.52 & 0.96 & 0.285$\pm$0.043 & 3509$^{+623}_{-460}$ & 3238$^{+465}_{-387}$ & 3710$\pm$297 & 5\\
\noalign{\smallskip}
\noalign{\smallskip}
\hline
\end{tabular}
\tablefoot{The magnitudes, goodness-of-fit (GOF), RUWE, and parallax ($\varpi$) from \textit{Gaia} DR3 are shown. The distances obtained from the inversion of the parallax are listed together with the geometric distances from \citet{2021AJ....161..147B}, and estimates from the published literature.\label{tab:gaia}\\$^a$GOF statistic for the astrometric solution. According to the \textit{Gaia} DR3 documentation, values $\gtrsim$3 indicate a poor fit to the data. $^b$Renormalized unit weight error (RUWE). Values around 1.0 are expected for sources well-fitted by a single-star model, while significantly higher values may indicate issues with the astrometric solution or the presence of binarity \citep[see, e.g.,][]{2022MNRAS.513.2437P,2024A&A...688A...1C}. Note that RUWE enters into the formula for the GOF. Therefore, if RUWE indicates a poor solution, it is probable GOF will as well.
\tablebib{(1)~\citet{1993AJ....106.1573K}; (2) \citet{2000A&A...353..952M}; (3) \citet{2014A&A...564A...1B}; (4) \citet{2005A&A...440..995S};
(5) \citet{1997ApJ...485..359G}.
}
}

\end{table*}

  \begin{figure*}
   \centering
   \includegraphics[width=0.98\columnwidth]{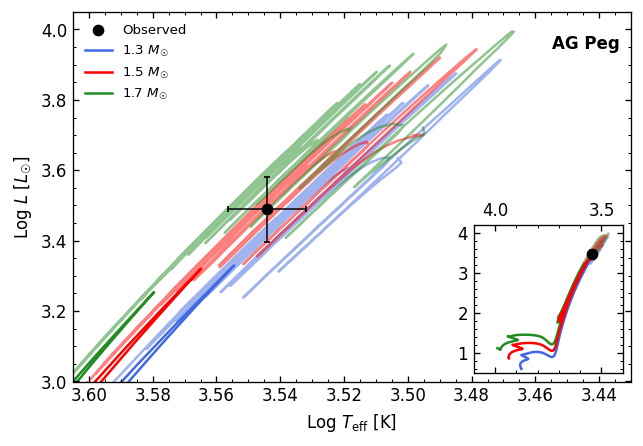}
   \includegraphics[width=0.99\columnwidth]{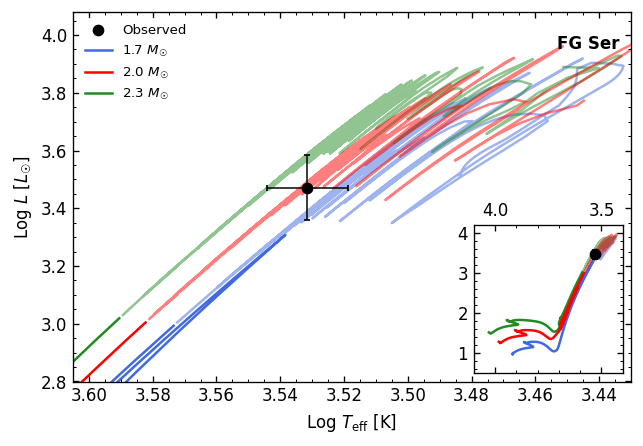}
   \includegraphics[width=0.98\columnwidth]{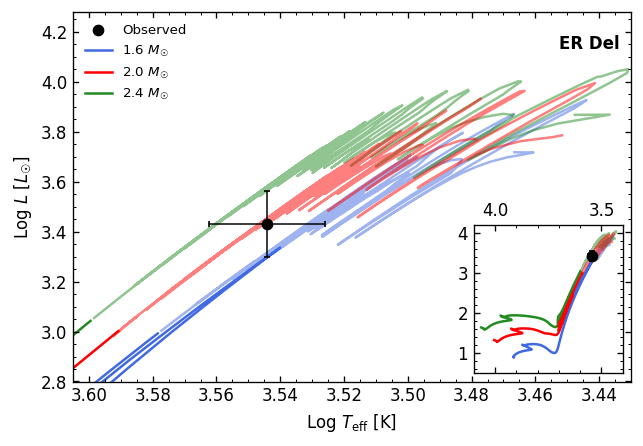}
   \includegraphics[width=0.98\columnwidth]{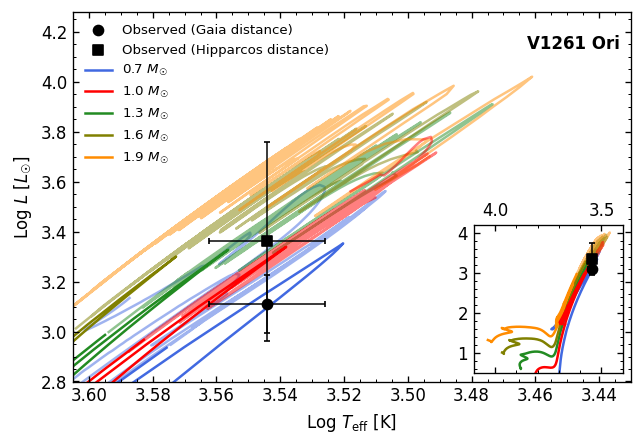}
   \includegraphics[width=0.98\columnwidth]{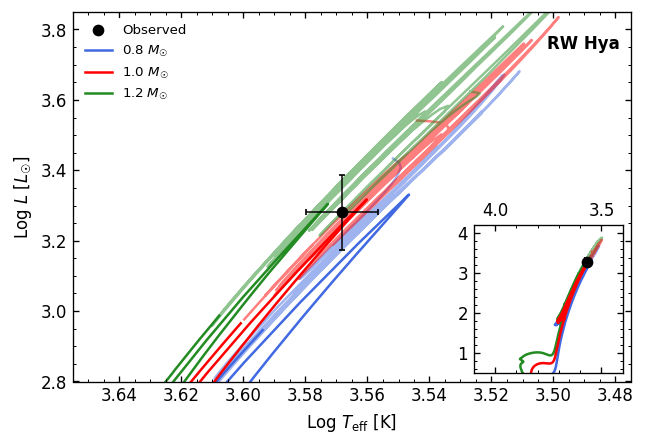}
   \includegraphics[width=0.98\columnwidth]{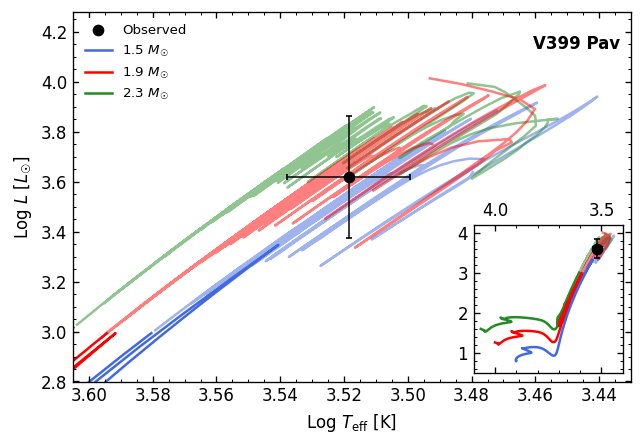}

      \caption{Position of target stars in the H-R diagram. For each star, three selected MESA evolutionary tracks \citep{2011ApJS..192....3P,2016ApJS..222....8D,2016ApJ...823..102C} are shown (in green, red, and blue), calculated for the specific metallicity of the star. The evolution before and during the first RGB is represented in darker shades, while the AGB evolution is depicted in lighter shades. The post-AGB evolution is omitted for clarity.
              }
         \label{fig:hr}
   \end{figure*}

\subsection{AG Peg}\label{sec:agpeg}
AG~Peg is one of the most well-known symbiotic stars, consisting of an M3 giant \citep{1999A&AS..137..473M} with a subsolar metallicity of [Fe/H] = -0.51 \citep{2023MNRAS.526..918G} and a shell-burning white dwarf on an 818-day orbit \citep{2000AJ....119.1375F}. It is also notable as the slowest symbiotic nova ever recorded, with an outburst that began in 1850, peaked around 1885, and returned to quiescence by 2000 \citep{1921AN....213...93L,1967SvA....11....8B,2017A&A...604A..48S}. Interestingly, AG~Peg experienced a Z And-type outburst 165 years after its initial nova flare-up \citep{2016MNRAS.462.4435T,2017A&A...604A..48S,2019CoSka..49..228M}. AG~Peg is also an X-ray source \citep{1997A&A...319..201M,2016MNRAS.461.3599R,2018MNRAS.481.5156Z}, likely due to shock-heated plasma from the collision of winds between the two components.

Given the very long duration of the nova outburst, the white dwarf in AG~Peg is likely of low mass. \citet{1993AJ....106.1573K} determined masses of $M_{\rm G}$ = 2.6$\pm$0.4 M$_\odot$ and $M_{\rm h}$ = 0.65$\pm$0.10 M$_\odot$ for an orbital inclination of 50\textdegree{} based on radial velocity analyses of the giant and the He II lines used as a proxy for the hot component. \citet{2010arXiv1011.5657M} estimated the mass of the hot component to be $M_{\rm h}$ = 0.46$\pm$0.10 M$_\odot$.

\citet{1993AJ....106.1573K} estimated a distance of 800 pc to AG~Peg from the calibration of absolute bolometric magnitudes, assuming the M3 giant has a luminosity class III. This distance estimate has been widely used in subsequent literature, for example by \citet{2017A&A...604A..48S} and \citet{2019ApJ...874..178S}. Although the \textit{Gaia} DR3 parallax signal-to-noise ratio of $\varpi$/$\sigma_\varpi \approx 21.5$ suggests that a simple parallax inversion could yield the distance to AG~Peg, the goodness-of-fit (GOF) statistic in the \textit{Gaia} archive is 9.56 — well above the threshold value of 3 — indicating a poor fit to the data. The renormalized unit weight error (RUWE) of 1.36, also above the recommended limit of 1.2, suggests potential issues with the astrometric solution.

Nevertheless, assuming the \textit{Gaia} distance ($\approx 1.3$ kpc) is accurate, we derived a linear radius for the cool component in AG~Peg of R$_{\rm G}$ = 151.3$^{+8.7}_{-7.9}$~R$_\odot$. For a giant mass of $M_{\rm G}$ = 2.6 M$_\odot$ and an inclination of 50\textdegree{} \citep{1993AJ....106.1573K}, the giant fills $\approx$0.56 of its Roche lobe. At a distance of 800 pc, the giant would fill even less, about 0.34. This conclusion remains valid even when considering the lower companion mass estimated by \citet{2010arXiv1011.5657M}, which corresponds to a red giant mass of 1.33 M$_\odot$ and a Roche-lobe filling factor of 0.72. Any reasonable change in the inclination would not significantly alter this result, leading to the overall conclusion that the red giant in AG~Peg is well within its Roche lobe, consistent with the non-detection of any ellipsoidal photometric variability \citep{2007BaltA..16...49R}.

The position of AG Peg in the Hertzsprung-Russell (H-R) diagram (Fig.~\ref{fig:hr}), adopting our interferometric radius, is consistent with the evolutionary tracks of a $\approx$1.5$\pm$0.2 M$_\odot$ star calculated for a metallicity of [Fe/H] = -0.51 \citep{2023MNRAS.526..918G}. The tracks are from the Modules for Experiments in Stellar Astrophysics \citep[MESA;][]{2011ApJS..192....3P}, obtained using the MIST Web Interpolator \citep{2016ApJS..222....8D,2016ApJ...823..102C}. Additionally, its position in the color-magnitude diagram (Fig.~\ref{fig:cmd}) based on the \textit{Gaia} $G$ and $RP$ magnitudes\footnote{It is important to keep in mind that nebular emission from the system could contribute to the flux observed in the \textit{Gaia} bands, making the object appear bluer and brighter.}, combined with the \textit{Gaia} distance and corrected for extinction of E$_{\rm (B-V)}$ = 0.06 mag that we obtained using the {\tt mwdust} code by \citet{2016ApJ...818..130B} from combined 3D dust maps of \citet{2003A&A...409..205D}, \citet{2006A&A...453..635M}, and \citet{2019ApJ...887...93G} suggests a similar mass for the giant. From the comparison with the evolutionary tracks, it seems that the giant is already on the AGB. The lower mass estimate of $\approx$1.5 M$_\odot$ aligns better with the hot component mass of \citet{2010arXiv1011.5657M} rather than the giant mass of 2.6 M$_\odot$ by \citet{1993AJ....106.1573K}. The mass would be even smaller if a lower distance were adopted.

Assuming an inclination of 50\textdegree{} and a giant mass of 1.5 M$_\odot$, the resulting hot component mass is 0.49 M$_\odot$, leading to a mass ratio of q$\approx$3. This is lower than the mass ratio obtained by \citet{1993AJ....106.1573K} from analyzing the radial velocities of the giant and using He II emission lines as a proxy for the motion of the hot component, thereby suggesting that these lines may overestimate the mass ratio.

Finally, we note that \citet{2007MNRAS.380.1053Z} claimed the rotation of the giant is synchronized with the orbit. However, for our derived radius, synchronization would require a projected rotational velocity, $v_{\rm rot} \sin{i}$, of approximately 9.36 km/s, whereas \citet{2003ASPC..303..113F} reported a much lower $v_{\rm rot} \sin{i}$ of only 4.5$\pm$1 km/s. This discrepancy suggests that the rotation of the giant is not yet synchronized with the orbital motion. \citet{2000AJ....119.1375F} also reported a slightly eccentric orbit, which is consistent with the idea that the system has not fully circularized or synchronized.

\subsection{FG Ser}
FG~Ser (AS 296) is a well-known symbiotic star first identified as an emission-line star by \citet{1950ApJ...112...72M} and later confirmed as a symbiotic source due to the presence of an M5 giant continuum and strong emission lines \citep{1970MmRAS..73..153W,1978MNRAS.184..601A,1999A&AS..137..473M}. FG~Ser underwent a symbiotic nova outburst beginning in 1988 \citep{1988IAUC.4622....2M,1989MNRAS.239..273M,1992AJ....104..262M}. The spectroscopic orbit presented by \citet{2000AJ....120.3255F} indicates a circular orbit with an orbital period of approximately 633 days. Similar periodicity is also observable in the photometric data \citep[see, e.g.,][]{2019CoSka..49...19S}. During the outburst, the light curves exhibited clear signs of eclipses.

Various distance estimates for FG~Ser have been presented in the literature, ranging from 0.9 kpc \citep[based on the comparison of the observed and tabulated absolute $K$ magnitude of an M5 III giant;][]{1992AJ....104..262M}, to 1.1$\pm$0.2 kpc obtained by \citet{2000A&A...353..952M} using the red giant radius of 105$\pm$15 R$_\odot$, calculated under the assumption of synchronized rotation from $v_{\rm rot} \sin{i}$ and the surface brightness relation for M giants by \citep{1999ESASP.427..397S}, up to 2.2 kpc \citep{1985PAZh...11...55T} based on a comparison of the optical and infrared brightness of FG~Ser with a sample of red giants. The \textit{Gaia} astrometry appears reliable for FG~Ser, given a GOF of 1.06, RUWE of 1.05, and $\varpi$/$\sigma_\varpi \approx 13.6$. 

FG~Ser has been observed on multiple occasions using VLTI/PIONIER and our analysis suggests that the angular diameter of this star varies (see Table \ref{tab:log}). However, there is no clear link to the orbital period of the system. A detailed analysis of this extensive dataset, along with new spectroscopic and photometric follow-up, is beyond the scope of this work and will be presented separately (Boffin et al., in prep.). Here, we calculated the average angular diameter across all epochs to draw conclusions. By adopting the distance obtained from the inversion of the \textit{Gaia} DR3 parallax, we determine the linear radius of the giant in FG~Ser to be R$_{\rm G}$ = 157.1$^{+14.8}_{-12.8}$ R$_\odot$.

Using this radius and an effective temperature of 3400 K \citep{2023MNRAS.526..918G}, we calculated the luminosity of the giant, L$_{\rm G}$~=~2959$^{+1019}_{-744}$ L$_\odot$, with uncertainties derived adopting a conservative temperature error of $\pm$100 K. Comparing this with evolutionary tracks for stars of various masses and metallicity of [Fe/H] = -0.08 \citep{2023MNRAS.526..918G}, we find that the position of FG~Ser in the H-R diagram is consistent with a mass of $M_{\rm G}$ = 2.0$\pm$0.3 M$_\odot$ of a star on the AGB (Fig.~\ref{fig:hr}). The color-magnitude diagram suggests a mass at the upper end of this range as well (Fig.~\ref{fig:cmd}). Assuming an inclination of $i = 90$\textdegree{}, the mass of the companion is calculated using the orbital elements (Table \ref{tab:target_stars}) to be $M_{\rm h}$ = 0.52$^{+0.04}_{-0.05}$ M$_\odot$.  
The corresponding Roche-lobe filling factor for the red giant ranges from 0.65 to 0.87. Lowering the inclination to 75\textdegree{} has minimal impact on the Roche-lobe filling factor and increases the mass of the companion only to $M_{\rm h}$ = 0.54 M$_\odot$.

Our findings are consistent with the conclusions of \citet{1995AJ....109.1740M}, who suggested that the giant in FG~Ser is still well within its Roche lobe, rather than with those of \citet{2014A&A...564A...1B}, who proposed that the giant is filling its Roche lobe. Their high Roche-lobe filling factor was based on the adopted giant mass of 1\,M$_\odot$, whereas our best estimate is $\approx$2\,M$_\odot$. Additionally, there is no evidence of ellipsoidal variability in the photometric light curves presented in the literature that would be expected if the giant were close to filling its Roche lobe and consequently tidally distorted.

Our calculated masses are slightly different but within the uncertainties of those reported by \citet{2000A&A...353..952M}, who found $M_{\rm G}$ = 1.7$\pm$0.7 M$_\odot$ and $M_{\rm h}$ = 0.6$\pm$0.15 M$_\odot$ using a similar methodology but with a smaller radius for the giant of 105$\pm$15 R$_\odot$. They adopted $v_{\rm rot} \sin{i}$ = 8$\pm$1 km/s and $\sin{i}$ = 1. Later, \citet{2007MNRAS.380.1053Z} reported $v_{\rm rot} \sin{i}$ = 9.8$\pm$1 km/s. Our calculated radius predicts a $v_{\rm rot}$ of approximately 12.55 km/s if the rotation of the star were synchronized with the orbital period, which does not appear to be the case. The observed $v_{\rm rot} \sin{i}$ values would require an inclination of 52\textdegree\ or less, which is inconsistent with the observed eclipses during the active stages of FG~Ser.

\subsection{ER Del}
ER Del was originally designated as an irregular variable star with a spectral type S5.5/2.5 \citep{1979ApJ...234..538A}, making it one of the two examples of S star AGB donors in the present sample. The symbiotic nature of this source was confirmed based on a strong UV continuum and emission lines with an ionization potential up to 47.5 eV reported by \citet{1989eprg.proc..371J}, detection of H$\alpha$ in emission by \citet{2002A&A...396..599V}, and the presence of strong X-ray emission \citep{2013A&A...559A...6L}. \citet{2024A&A...683A..84M} did not detect any flickering variability with the Transiting Exoplanet Survey Satellite (TESS) data.

With an orbital period of P$_{\rm orb}$ = 2081 days \citep{2019A&A...626A.127J}, this binary has the longest period among the systems studied here and belongs to a relatively small group of confirmed symbiotic stars with periods exceeding 2000 days (excluding D-type symbiotics, which have orbital periods spanning decades). A slightly eccentric orbit with e = 0.28 has been reported \citep{2012BaltA..21...39J,2019A&A...626A.127J,2014A&A...564A...1B}. Based on typical masses for components in systems similar to ER Del, \citet{2014A&A...564A...1B} estimated the distance to be in the range of 1750 -- 2600 pc, consistent with the \textit{Gaia} parallax-based distance of 2070 pc. The \textit{Gaia} astrometric solution for ER Del appears reliable, with a GOF of 1.24, RUWE of 1.04, and $\varpi$/$\sigma_\varpi \approx 19.3$.

For this distance, the corresponding radius of the giant star, based on our interferometric angular diameter, is R$_{\rm G}$ = 141.5$^{+9.9}_{-8.9}$~R$_\odot$. \citet{2014A&A...564A...1B} assumed a typical white dwarf mass in a post-mass transfer system to be 0.7\,M$_\odot$, estimating the mass of the giant to be M$_{\rm G}$ = 2.0 -- 3.5\,M$_\odot$ for an inclination of i = 50\textdegree{} -- 90\textdegree{}. The position of ER Del in the H-R diagram (Fig. \ref{fig:hr}), compared with MESA evolutionary tracks of stars with a metallicity of [Fe/H] = -0.18 dex — the median metallicity for S-type symbiotic stars \citep{2019RNAAS...3...28M,2023MNRAS.526..918G} — suggests a giant mass of about 2.0$\pm$0.4 M$_\odot$. With this mass, an inclination of 50\textdegree{} results in a rather large companion mass of 1.15 M$_\odot$, while i = 90\textdegree{} yields a companion mass of 0.82 M$_\odot$. A larger giant mass would imply a companion mass exceeding typical values for such systems, while a lower giant mass would place the star on the red giant branch (RGB) rather than the AGB, which is inconsistent with the S-star nature of the cool component in ER Del. Regardless, the giant is far from filling its Roche lobe under any reasonable assumptions. For example, for M$_{\rm G}$ = 2.0 M$_\odot$ and i = 50\textdegree{} -- 90\textdegree{}, the Roche-lobe filling factor is approximately 0.33.

\subsection{V1261 Ori}\label{sec:ori}
V1261 Ori (also HD 35155) is the second symbiotic star with an S star AGB donor in our sample. Its cool component is classified as an S4.1 giant \citep{1954ApJ...120..484K,1996A&A...306..467J} with a subsolar metallicity, [Fe/H] = -0.53 $\pm$ 0.35 dex \citep{2016A&A...585A..64S}. The star was identified as symbiotic due to strong emission lines with ionization potential up to 77.5 eV in UV \citep{1991ApJ...383..842A} and X-ray emission from the source \citep{1996A&A...306..467J}. \citet{2019AN....340..598M} classified its X-ray spectrum as of the $\delta$-type symbiotics. Based on eclipses observed in UV and optical \citep{1991ApJ...383..842A,1992IBVS.3730....1J}, the system has been classified as an eclipsing binary, suggesting an inclination close to 90\textdegree{}.

\begin{figure}
   \centering
   \includegraphics[width=\columnwidth]{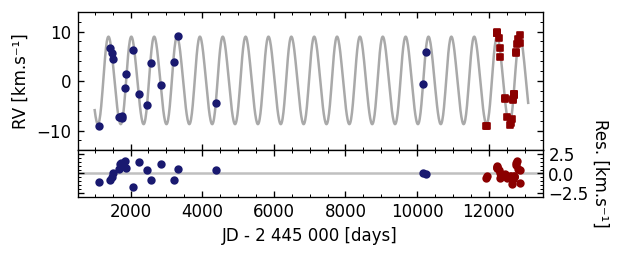}
\includegraphics[width=0.49\columnwidth]{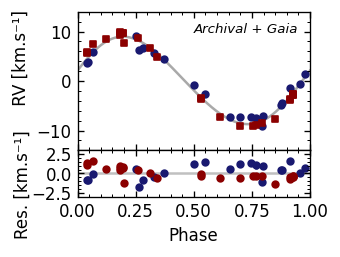}
\includegraphics[width=0.49\columnwidth]{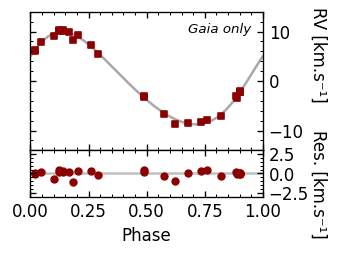}

      \caption{Orbit of V1261 Ori. Upper panel: Orbital solution based on archival data \citep[blue data points from CORAVEL and Mercator/HERMES;][]{2014A&A...564A...1B} and radial velocities (shown in dark red) published in the \textit{Gaia} Focused Product Release \citep{2023A&A...680A..36G}. Lower panels: Phased radial velocities based on the combined archival and \textit{Gaia} data (left) and \textit{Gaia} data only (right). See Table \ref{tab:v1261} for the obtained orbital elements.
              }
         \label{fig:orbit}
   \end{figure}

The optical light curve reveals signs of ellipsoidal variability \citep{2013AcA....63..405G,2014A&A...564A...1B}, and semi-regular pulsations with a period around 56 days are detected. Orbital elements were published by \citet{1992A&A...260..115J,1998A&A...332..877J} and \citet{2014A&A...564A...1B}, with additional radial velocities now available from the \textit{Gaia} Focused Product Release \citep{2023A&A...680A..36G}. Using {\tt radial} code\footnote{\url{https://github.com/ladsantos/radial}}, we analyzed these data to refine the orbital elements. The results are shown in Table \ref{tab:v1261} and the orbit in Fig. \ref{fig:orbit}. We analyzed the archival and \textit{Gaia} data separately and together. It is clear from this exercise that the limited time span of \textit{Gaia} data compared to the orbital period of 638 days leads to a slight period discrepancy for V1261 Ori. This should be considered when analyzing binaries of similar orbital periods with the \textit{Gaia} DR3 data \citep[see also][]{2024A&A...682A...7B}. Interestingly, \textit{Gaia} radial velocities suggest a slightly higher amplitude than the archival data, resulting in a slightly larger mass function. For further analysis, we adopted values from the combined data, closely matching those of \citet{2014A&A...564A...1B}.

\begin{table}[]
\centering
\small
\caption{Revised orbital elements of V1261 Ori.\label{tab:v1261}}%
\begin{tabular}{llll}
\hline
\noalign{\smallskip}
 & Archival & \textit{Gaia} & Arch. + \textit{Gaia} \\
 \hline
  \noalign{\smallskip}
$P$ {[}d{]} & 638.23$\pm$0.15 & 611.55$\pm$4.09 & 638.09$\pm$0.09 \\
$e$ & 0.103$\pm$0.007 & 0.135$\pm$0.007 & 0.061$\pm$0.003 \\
$T_0$ 2 453.. {[}d{]} & 267.26$\pm$11.45 & 478.93$\pm$32.20 & 278.82$\pm$9.56 \\
$K$ {[}km/s{]} & 8.14$\pm$0.09 & 9.35$\pm$0.06 & 8.84$\pm$0.05 \\
$\omega$ {[}\textdegree{]} & -85.6$\pm$7.1 & -61.9$\pm$4.2 & -76.9$\pm$5.6 \\
$f$(m) {[}M$_\odot${]} & 0.035089  &  0.050368 &  0.045403\\
O-C {[}km/s{]} & 0.55 & 0.29 & 0.73\\
\noalign{\smallskip}
\hline
\end{tabular}
\end{table}

The only distance estimate available in the literature, 510$^{+375}_{-215}$ pc \citep{2014A&A...564A...1B}, is based on \textsc{Hipparcos} data \citep{2000A&AS..145..161P}, while \textit{Gaia} suggests a slightly lower distance of 382$^{+19}_{-17}$ pc. However, \textit{Gaia} astrometry for this source seems to be unreliable, with a GOF of 69.75 and RUWE of 4.77, despite a parallax signal-to-noise ratio of $\varpi$/$\sigma_\varpi \approx 20.9$.

Using our measured angular diameter of the giant component of V1261 Ori and the \textit{Gaia}-based distance of 382 pc, the radius of the giant is R$_{\rm G}$ = 97.8$^{+5.0}_{-4.5}$ R$_\odot$. For a \textsc{Hipparcos}-based distance of 510 pc, we find R$_{\rm G}$ = 130.6$^{+96.3}_{-55.2}$ R$_\odot$. Assuming an inclination of 90\textdegree{} and our calculated mass function, the giant would only fill its Roche lobe for implausibly low masses of M$_{\rm G}$ < 0.27 M$_\odot$ or < 0.59 M$_\odot$. Given more likely giant masses of 1.3 -- 1.8 M$_\odot$ \citep{1992IBVS.3730....1J}, the star would fill only 50 -- 56\% or 66 -- 75\% of its Roche lobe with either distance estimate. Any change in the inclination would not significantly influence this conclusion. This finding contradicts observations of ellipsoidal variability, which suggest a Roche-lobe filling factor close to unity, as the amplitude of the variability depends on the third power of the Roche-lobe filling factor \citep{1990AJ....100..554H,2004AcA....54..347S}.

The estimated radii present further discrepancies in the H-R diagram when compared with evolutionary tracks for stars with [Fe/H] = -0.53 dex (Fig. \ref{fig:hr}). At 98 R$_\odot$ (for \textit{Gaia}-based distance), the star lie near the tip of the RGB of a 0.7 M$_\odot$ star. The \textsc{Hipparcos} distance, though with very large uncertainties, would place it among evolutionary tracks for stars with 0.6 -- 2.0 M$_\odot$. \textit{Gaia} CMD would lead to similar conclusion (Fig. \ref{fig:cmd}). Adjusting the metallicity to -0.18 dex, the upper limit from \citet{2016A&A...585A..64S}, predicts a more realistic mass range of $\approx$1.0 -- 1.3 M$_\odot$ for the \textit{Gaia} distance (Fig. \ref{fig:v1261ori_alternative}).

Assuming now the star fully fills its Roche lobe and has a mass of 1.3 -- 1.8 M$_\odot$ \citep{1992IBVS.3730....1J}, the estimated Roche-lobe radius is 174.6 -- 197.2 R$\odot$. This implies a distance of 682 -- 770 pc ($\varpi$ = 1.30 -- 1.47 mas), within \textsc{Hipparcos} uncertainties and approximately twice the \textit{Gaia} distance. Such discrepancies may result from the binary nature of V1261 Ori, as the \textit{Gaia} astrometry should be considered unreliable based on the GOF and RUWE values. Given that the binary motion adds an additional component to the center-of-light motion on the sky, single-star models may have overestimated the parallax. However, the discrepancy seems to be too large in this case to account solely for the binarity (see the detailed discussion in Appendix \ref{sec:binarity}). \textit{Gaia} DR4 should significantly improve the accuracy, especially when the time-series data will be available for the community for analysis. Even if we assume that the distance is underestimated, one inconsistency remains: if the giant fully fills its Roche lobe, synchronous rotation and orbit circularization would be expected, yet neither are observed \citep[see Table \ref{tab:v1261} and][]{2014A&A...564A...1B}.

Alternatively, the Roche-lobe may be lower than the canonical value due to stellar wind or pulsations, meaning the giant could be filling its Roche lobe even though it is smaller than what would be predicted by Eggleton's formula \citep{1983ApJ...268..368E}. \citet{2009A&A...507..891D} explored such a possibility, finding Roche radii could decrease by a factor of $\approx$2, depending on stellar characteristics. However, their results also indicated that such a reduction in size does not affect the stability of the mass transfer. That would imply that V1261 Ori and, eventually, also other similar systems should be in the unstable mass transfer regime. It also might be that a slow, dense wind is causing the observed ellipsoidal variability. We plan to investigate these possibilities further using new VLTI/GRAVITY and VLTI/PIONIER observations. If the giant is tidally distorted, such a deformation should be detectable if observations are conducted at quadrature; a similar approach was successfully applied in the case of HD 352 \citep{2024A&A...692A.218M}.

\subsection{RW Hya}\label{sec:rwhya}
RW Hya is one of the earliest identified symbiotic stars. Its peculiar spectrum was first noted by \citet{1932PASP...44...56M} and \citet{1933ApJ....78...87M}. The spectrum of RW Hya is characterized by strong emission lines with ionization potentials up to 54.4 eV (He II) and a continuum consistent with an M2 giant \citep{1999A&AS..137..473M}. High-resolution spectral analyses suggest that the giant is metal-poor, with [Fe/H] $\approx -0.7$ \citep{2014MNRAS.440.3016M,2016MNRAS.455.1282G,2017ApJ...841...50P}. To date, no outburst has been recorded for this star \citep{1995AJ....110..391K,2005A&A...440..995S}. Spectroscopic observations by \citet{1995AJ....110..391K} and \citet{1996A&A...306..477S} have provided well-constrained orbital elements, indicating an orbital period of $\approx$370 days and a circular orbit. The system is also known to exhibit eclipses in the UV \citep{1995AJ....110..391K,1996A&A...306..477S} and photometric variability with half the orbital period in the infrared, explained as a consequence of a tidally deformed giant \citep{2014ASPC..490..367O}.

Several distance estimates for RW Hya have been proposed in the literature. \citet{1996A&A...306..477S} derived a distance of 670$\pm$100 pc from observed infrared magnitudes and the luminosity, assuming a radius of 58.5$\pm$8 R$_\odot$, based on $v_{\rm rot}$ and the assumption of synchronized rotation. \citet{2005A&A...440..995S} used the same radius combined with spectral energy distribution modeling to estimate a slightly higher distance of 820$\pm$112 pc. In contrast, \citet{1980ApJ...240..114K} suggested a larger distance of 1300 pc by comparing the observed $V$ magnitude with the tabulated values for M2 III stars. While the \textit{Gaia} DR3 parallax for RW Hya has a high signal-to-noise ratio ($\varpi$/$\sigma_\varpi \approx 18.5$), the GOF (6.79) and RUWE (1.27) suggest the astrometric solution may be unreliable.

Adopting the \textit{Gaia} parallax-based distance of 1543 pc, we calculated a linear radius for the giant, R$_{\rm G}$ = 106.5$^{+7.8}_{-7.0}$ R$_\odot$. This radius is nearly twice as large as the one obtained by \citet{1996A&A...306..477S}, who based their estimate on $v_{\rm rot}$ = 8$\pm$1 km/s, assuming synchronous rotation. It is worth noting that \citet{2007MNRAS.380.1053Z} reported slightly lower $v_{\rm rot} \sin{i}$ values of 6.2$\pm$1 km/s and 7.1$\pm$1.5 km/s, corresponding to even smaller radii of 45.4 and 52.0 R$_\odot$ (assuming $i$ = 90\textdegree{}). The value 5.0$\pm$1 km/s reported by \citet{2003ASPC..303..113F} predicts the radius of 36.6 R$_\odot$ only.

In contrast, \citet{2014ASPC..490..367O} obtained significantly larger radii of 144.5 and 111.9 R$_\odot$ (with and without a mass ratio constraint, respectively) from modeling the ellipsoidal variability in infrared light curves. The latter value aligns with our interferometric radius within the errors. Their best fit was achieved with an inclination of 80\textdegree{} and a mass ratio of q = 3.2 ($M_{\rm G}$ = 1.6 M$_\odot$, $M_{\rm h}$ = 0.5 M$_\odot$; exactly the same as the one obtained by \citeauthor{1996A&A...306..477S} \citeyear{1996A&A...306..477S} when they compared the position of RW Hya in the H-R diagram with evolutionary tracks). The larger value of radius was derived by constraining the mass ratio using radial velocity measurements of the He II 1640 \AA{} and He II 4686 \AA{} lines as proxies for the motion of the hot component. This approach yielded masses $M_{\rm G}$ = 3.4 M$_\odot$ and $M_{\rm h}$ = 0.8 M$_\odot$, and an inclination of 75\textdegree{}.

Comparing the position of RW Hya in the H-R diagram (Fig.~\ref{fig:hr}), using our interferometric radius and its position in the color-magnitude diagram (Fig.~\ref{fig:cmd}) based on the \textit{Gaia} distance and reddening from the dust map (E$_{\rm (B-V)}$ = 0.06 mag), with the MESA evolutionary tracks calculated for a metallicity of [Fe/H] = -0.76 \citep{2014MNRAS.440.3016M}, supports the conclusion that the giant has a low mass ($\approx$1.0 - 1.3 M$_\odot$) and is either at the tip of the RGB or already on the AGB. The evolutionary tracks for the masses inferred by \citet{2014ASPC..490..367O} from infrared light curve analysis lie far from our determined position of RW Hya. The slightly larger mass of 1.6 M$_\odot$ inferred by \citet{1996A&A...306..477S} was derived in a similar manner to ours; however, they compared the position of RW Hya with tracks assuming super-solar metallicities and adopted a smaller radius, leading to a lower luminosity.

Assuming an inclination of 90\textdegree{}, the corresponding mass of the companion would be relatively low for our estimated mass range of the giant, around $\approx$0.4 M$_\odot$, resulting in a mass ratio of 2.7 -- 3.0. As in the case of AG Peg (Sect. \ref{sec:agpeg}), the mass ratio obtained from the radial velocities of the giant and those inferred from the He II emission lines — assuming they trace the hot component's motion — is significantly larger, q = 4.8$\pm$0.6 \citep{2014ASPC..490..367O}.

In any case, there is clearly a discrepancy between the radii (and resulting parameters) obtained through different methods. If we adopt our interferometric radius, the orbital elements from Table \ref{tab:target_stars}, and assume an inclination of 90\textdegree{}, the giant would fill its Roche lobe only if $M_{\rm G}$ $\lesssim$ 0.86 M$_\odot$. Decreasing the inclination to 70\textdegree{} raises this limit by just 0.02 M$_\odot$. For masses of $M_{\rm G}$ = 1.6 and 3.4 M$_\odot$ \citep{2014ASPC..490..367O}, the corresponding Roche lobe filling factors would be $\approx$ 0.80 and 0.61, respectively.

Conversely, adopting a lower distance estimate, such as the one by \citet{2005A&A...440..995S}, results in a smaller radius of 56.6 R$_\odot$, (not surprisingly closely matches the value used by the authors to calculate that distance), consistent with the observed $v_{\rm rot}$ under the assumption of synchronous rotation. However, for a star with this smaller radius to fill its Roche lobe, it would require an implausibly low mass of $M_{\rm G}$ $\lesssim$ 0.16 M$_\odot$. If we again use the masses obtained by \citet{2014ASPC..490..367O} from IR light curve modeling, the Roche lobe filling factors would be as low as 0.42 and 0.33, respectively.

As in the case of V1261 Ori (Sect. \ref{sec:ori}), it might be that even the \textit{Gaia} distance, largest from the available distance estimates, is underestimated. If this is the case, our calculated radii would be larger, and the giant could be filling its Roche lobe within a more reasonable range of masses. For example, for masses of $M_{\rm G}$ = 1.3 M$_\odot$ and 1.6 M$_\odot$, the giant would fill its Roche-lobe if in the distance of 1800 pc and 1946 pc ($\varpi$ = 0.56 mas and 0.51 mas), respectively. This is larger but close to the \textit{Gaia} distance, and such a discrepancy might be due to the binarity \citep[see Appendix \ref{sec:binarity} and][]{2020MNRAS.495..321P}. In such a case, however, the rotation would definitely not be synchronized with the orbital motion. Alternatively, as discussed for V1261 Ori, Eggleton's formula might not correctly predict the Roche-lobe size or the wind may account for some of the ellipsoidal variability.

\subsection{V399 Pav}
V399 Pav (Hen 3-1761) is a relatively poorly studied symbiotic system first identified as a star with a peculiar spectrum by \citet{1951BHarO.920...32M}. It was later classified as a symbiotic star based on the presence of emission lines with ionization potentials up to 99 eV in its spectra \citep{1954Obs....74..258T,1998A&AS..132..281B}, along with the continuum of an M5-type star \citep{1997ApJ...485..359G,1999A&AS..137..473M}. The giant component of the system has a subsolar metallicity ([Fe/H] = -0.25 dex), as determined from a high-resolution infrared spectrum by \citet{2016MNRAS.455.1282G}.

\begin{figure}
   \centering
   \includegraphics[width=0.98\columnwidth]{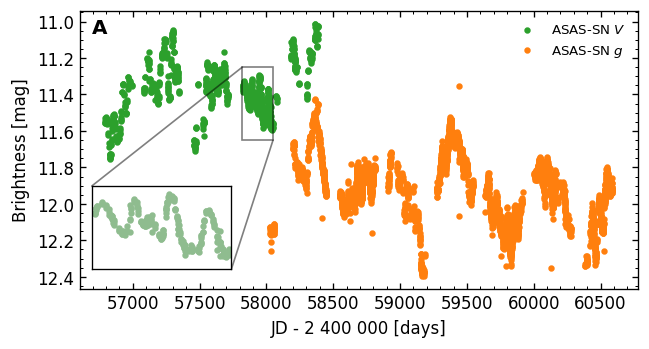}
\includegraphics[width=\columnwidth]{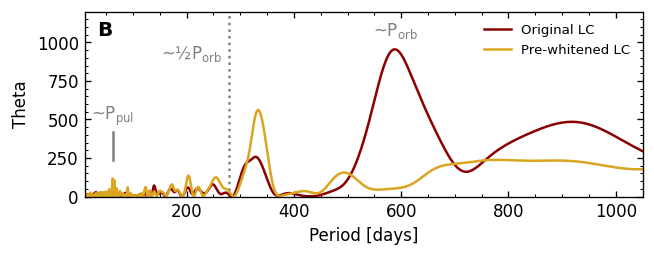}

      \caption{ASAS-SN light curve of V399 Pav. \textbf{A}: Light curves in $V$ and $g$ bands (in green and orange, respectively), with a zoomed-in section highlighting the pulsations. \textbf{B}: Results of the period analysis for the combined $V$- and $g$-band light curves (dark red) and for the pre-whitened light curve after removing the dominant period (dark yellow). 
              }
         \label{fig:v399pav_asassn}
   \end{figure}

An outburst of V399 Pav was reported in 1998–1999 by \citet{2006BAAA...49..132B}, who also provided the only available spectroscopic orbit for the system. The orbital period of 560 days was also detected in photometric data by \citet{2013AcA....63..405G}, along with an additional period of 63 days attributed to the pulsation of the cool giant. Both periods are clearly observable in more recent All Sky Automated Survey for SuperNovae (ASAS-SN) observations \citep[Fig.~\ref{fig:v399pav_asassn};][]{2014ApJ...788...48S,2017PASP..129j4502K}, although the variability seems to be rather complex.

\citet{1998A&AS..132..281B} estimated a distance of 2.2 kpc, assuming the cool component is a typical M5 giant. Alternatively, \citet{1997ApJ...485..359G} derived a distance of 3.7 kpc using a different calibration of absolute $K$ magnitudes. The distance derived from the \textit{Gaia} parallax (3.5 kpc) aligns well with the latter estimate within the uncertainties. The astrometric solution appears to be reasonably reliable, given the RUWE, GOF, and a parallax signal-to-noise ratio of $\varpi$/$\sigma_\varpi \approx 6.6$ (see Table \ref{tab:gaia}), although the distance errors are relatively large, as V399 Pav is the most distant and faintest star in this study.

Interferometric observations for V399 Pav are available from two epochs. Although the observations from October 2014 are of lower quality, the two epochs yield a similar angular diameter for the star (Table \ref{tab:log}). By adopting the distance inferred from the inversion of the \textit{Gaia} parallax and averaging the two angular diameters, the linear radius of the cool component in V399 Pav is calculated to be R$_{\rm G}$ = 197.4$^{+48.4}_{-35.7}$ R$_\odot$.

The inclination of the orbit of V399 Pav remains unknown, as do the masses of its individual components. However, the relatively large amplitude of the photometric variability suggests that the inclination is unlikely to be very low. For inclinations of 30\textdegree{}, 50\textdegree{}, and 90\textdegree{}, the giant in V399 Pav would fully fill its Roche lobe if its mass is below approximately $M_{\rm G}$ $\lesssim$ 2.63, 2.29, and 2.10 M$_\odot$, respectively. The median mass of symbiotic giants is about 1.5 M$_\odot$ \citep{2007BaltA..16....1M, 2010arXiv1011.5657M, 2019RNAAS...3...28M}.

However, it is important to note that the first scenario (i.e., $i$ = 30\textdegree{}) would require a companion mass of approximately 1.43 M$_\odot$. Even at an inclination of 50\textdegree{}, the white dwarf would be more massive than 1.0 M$_\odot$ if the mass of the giant exceeds 3.5 M$_\odot$. Most symbiotic white dwarfs have masses ranging between 0.4 and 0.8 M$_\odot$, with a median of 0.6 M$_\odot$ \citep{2007BaltA..16....1M, 2010arXiv1011.5657M, 2019RNAAS...3...28M}. Therefore, a lower giant mass and, consequently, a larger Roche lobe filling fraction seems to be preferred. 

The position of V399 Pav in the H-R diagram (Fig.~\ref{fig:hr}), when compared with MESA evolutionary tracks calculated for a metallicity of -0.25 dex \citep{2016MNRAS.455.1282G}, suggests that the mass of the giant is in the range of 1.5 -- 2.3 M$_\odot$ and that it may already be on the AGB, possibly experiencing thermal pulses. Such a giant mass implies larger Roche lobe filling factors for any reasonable inclination, but the uncertainties are too large to draw a firm conclusion. Unfortunately, infrared photometry is unavailable to study the presence of ellipsoidal variability. The ASAS-SN light curve is complex (Fig.~\ref{fig:v399pav_asassn}A), making it challenging to evaluate whether the giant fills its Roche lobe. The dominant period detected in the data using the ``date compensated'' discrete Fourier transform of \citet{1981AJ.....86..619F} implemented in Peranso software \citep{2016AN....337..239P} is 587.9$\pm$7.7 days (Fig.~\ref{fig:v399pav_asassn}B), close to the reported orbital period of the system. After pre-whitening the light curve with this period, residuals with a shorter period of 333.3$\pm$3.0 days become detectable (Fig.~\ref{fig:v399pav_asassn}B). This period does not seem to exactly coincide with the half of the orbital period ($\approx$ 280 days) expected from the ellipsoidal variability.

  

\section{Conclusions}

In this study we analyzed archival VLTI/PIONIER interferometric observations of six well-known symbiotic stars: AG Peg, FG Ser, ER Del, V1261 Ori, RW Hya, and V399 Pav. We fitted their interferometric observables — primarily squared visibilities, as the observed closure phases did not indicate any asymmetries — using limb-darkened disk models to determine their angular diameters (which range from 0.5 to 2.4 mas). Using \textit{Gaia} DR3-based distances, we calculated their linear radii and, subsequently, discussed their Roche-lobe filling factors. 

For all the studied systems, with the possible exception of the less certain case of V399 Pav, the giants appear to reside well within their canonical Roche lobes. This finding is particularly surprising for V1261 Ori and RW Hya, as both systems exhibit ellipsoidal variability attributed in the literature to tidally distorted giants close to filling their Roche lobes. For V1261 Ori, our detailed analysis of the effects of binarity on \textit{Gaia} astrometry suggests that the discrepancy cannot be explained solely by an underestimated \textit{Gaia} distance. This implies that additional mechanisms must account for the apparent inconsistency between the sizes of the giants relative to their Roche lobes and the observed ellipsoidal variability. Possible explanations include modifications to the canonical Roche-lobe size, as proposed in the literature, or an alternative source for the variability, such as a slow, dense wind filling the Roche lobe rather than the giant itself.

A comparison of the positions of studied symbiotic giants with evolutionary tracks indicates that most if not all are luminous stars already on the AGB. However, within the uncertainties, it cannot be ruled out that at least RW Hya is at the tip of the first RGB. This suggests that deriving radii, masses, or distances for these objects based on calibrations for normal luminosity class III giants, as is often done in the literature, may not be valid. Additionally, it appears that the rotation of these giants is not yet (or no longer) synchronized with their orbital periods, meaning that radius estimates based on this assumption could significantly underestimate the true values.

Finally, the mass ratios between the giants and their white dwarf companions derived from our analysis are smaller than those previously inferred from the radial velocities of the giants and the He II emission lines used as a proxy for the compact component. This discrepancy, observed in both AG Peg and RW Hya, for which an estimate based on He II lines is available, suggests that the approach may systematically overestimate the mass ratios and deserves further investigation.

\begin{acknowledgements}
We thank the anonymous reviewer for the careful reading of the manuscript. The research of J.M. was supported by the Czech Science Foundation (GACR) project no. 24-10608O and by the Spanish Ministry of Science and Innovation with the grant no. PID2023-146453NB-100 (PLAtoSOnG). J.M. acknowledges the support received through the on-the-job training programme at the European Southern Observatory (ESO), funded by the Ministry of Education, Youth and Sports of Czechia (MEYS).
\end{acknowledgements}

%
\bibliographystyle{aa} 
\bibliography{bibliography} 
%

\begin{appendix} 

\section{Influence of binarity on the \textit{Gaia} astrometry}\label{sec:binarity}

For several stars in this study, the GOF statistics and RUWE values indicate a poor fit to the \textit{Gaia} astrometric data, implying unreliable parallaxes and inferred distances. The astrometric solutions in \textit{Gaia} DR3 were based on a single-star model, which does not account for binary motion. Depending on the binary parameters (orbital period, mass ratio, luminosity ratio, etc.), the offset between the motion of the center of mass of the system and the center of light observed by \textit{Gaia} can introduce significant additional motion affecting the astrometry. \citet{2020MNRAS.495..321P} demonstrated that this effect on parallax is most pronounced for binaries with orbital periods close to one year (with deviations of up to 1 mas in more extreme cases). In systems with much longer orbital periods, the binarity primarily influences proper motions.

\subsection{Case of V1261 Ori}

\begin{figure}
   \centering
   \includegraphics[width=0.9\columnwidth]{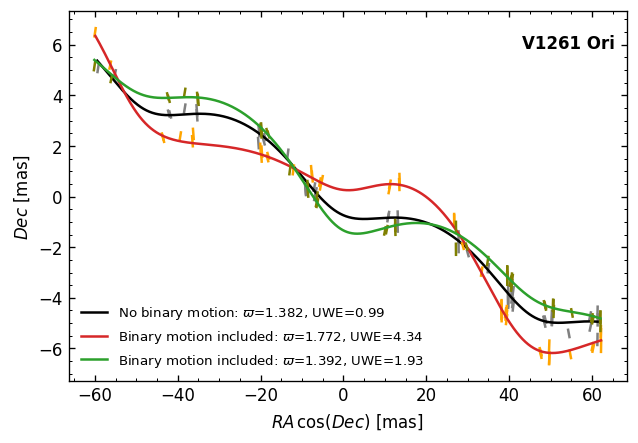}

      \caption{Examples of astrometric tracks of V1261 Ori. The black line represents the track of a single star at the position of V1261 Ori, with the same brightness, proper motions, and distance. Red and green are two examples of tracks with binarity included. Parallax and unit weight error values from the single-star model fit are indicated in the plot for each track.      
              }
         \label{fig:v1261ori_tracks}
   \end{figure}

The investigation of this issue is particularly relevant for V1261 Ori and RW Hya as their Roche-lobe filling factors estimated for the \textit{Gaia} parallax-based distances are not consistent with the observed ellipsoidal variability. For V1261 Ori, the \textit{Gaia} DR3 parallax is 2.617$\pm$0.125 mas (corresponding to the distance of 382 pc). At this distance, the giant would fill only about half of its Roche lobe (Sect. \ref{sec:ori}), contrary to observations of ellipsoidal variability. Fully filling the Roche lobe requires a distance of $\approx$680 -- 770 pc ($\varpi$ = 1.30 -- 1.47 mas). According to the analysis presented by \citet{2020MNRAS.495..321P}, the difference between the predicted parallax value and the \textit{Gaia} DR3 parallax appears too large to be explained solely by binary motion. Their study provided a general assessment of binary effects on astrometry for a broad parameter space. To better understand the specific case of V1261 Ori, we estimated the impact of binarity on \textit{Gaia} astrometry using tailored simulations. Using the actual \textit{Gaia} DR3 scanning law obtained via the {\tt gaiascanlaw} code\footnote{\url{https://github.com/zpenoyre/gaiascanlaw}} and the {\tt astromet} package\footnote{\url{https://github.com/zpenoyre/astromet.py}}, we generated synthetic astrometric measurements of V1261 Ori. The simulations incorporated realistic astrometric uncertainties based on the observed \textit{Gaia} G-band brightness of the target. We fixed the proper motions to the values from \textit{Gaia} DR3, and adopted the orbital parameters (period, eccentricity) from Table \ref{tab:v1261}. The semimajor axis of the binary (a = 1.87 au), mass ratio (q = $M_{\rm G}$/$M_{\rm h}$ = 2.6), and parallax ($\varpi$ = 1.38 mas) were set based on the assumption that the giant is filling its Roche lobe. Luminosity ratio was set to 0.01. If the companion (the hot component and/or nebula) actually contributes more to the observed flux (which easily can be the case of some symbiotic with strong nebular emission), the influence of binarity on astrometry would be actually lower than estimated here. The orbital orientation and the time of the periastron passage were randomized over 100\,000 iterations. Examples of the resulting astrometric tracks are shown in Fig. \ref{fig:v1261ori_tracks}.

     \begin{figure}
   \centering
   \includegraphics[width=0.9\columnwidth]{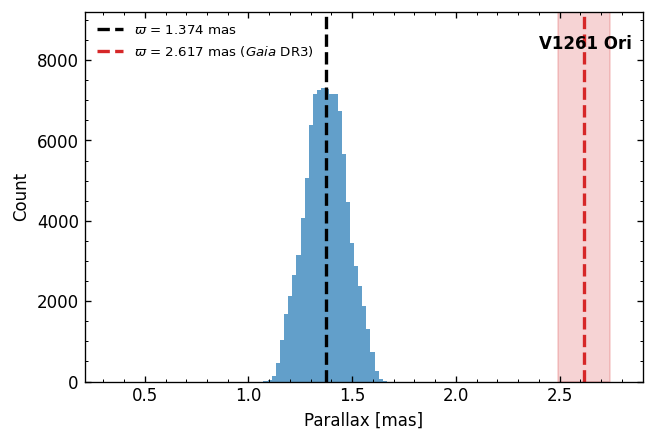}

      \caption{
      Parallaxes from our simulations of astrometric data of V1261 Ori. In each of the 100\,000 iterations, the orbital orientation and time of periastron passage were varied randomly. The input parallax is indicated by the dashed black line, while the parallax from the \textit{Gaia} DR3 catalog is marked by the dashed red line.}

         \label{fig:v1261ori_astrometry}
   \end{figure}

The simulated data, which included parallax and binary motion, were then fitted with a single-star model using the {\tt astromet} code. The resulting parallax values and unit weight error obtained from the fitting (Fig. \ref{fig:v1261ori_astrometry}) clearly indicate that binary motion can affect the inferred parallax of V1261 Ori. However, the amplitude of these deviations in the specific case of this target is smaller than the upper limit found for this orbital period by \citet{2020MNRAS.495..321P}. This is because the center of light is located closer to the more massive component (the giant), resulting in a smaller offset from the center of mass of the system compared to a scenario with an inverted mass ratio. The deviations would be more pronounced if the more luminous component were significantly less massive, as the radii of the individual orbits are inversely proportional to the stellar masses.

Importantly, if V1261 Ori were at a distance where the giant would fill its Roche lobe (as determined using Eggleton’s formula), the binary motion alone could not account for the discrepancy between the expected parallax at this distance and the \textit{Gaia} DR3 value (Fig. \ref{fig:v1261ori_astrometry}). This indicates that the inconsistency between the Roche-lobe filling factor and the observed ellipsoidal variability for V1261 Ori cannot be explained solely by an underestimated distance.

Additional factors have to be considered. Evolved giant and supergiant stars often exhibit a small number of large-scale convective cells on their surfaces \citep[see, e.g., the VLTI/PIONIER interferometric observations of {\ensuremath{\pi}}$^{1}$ Gruis;][]{2018Natur.553..310P}, which can introduce additional photocenter displacements. Recent hydrodynamic simulations of \citet{2024A&A...690A.125B} suggest that these displacements can range from 4 to 13\% of the stellar radius \citep[see also][]{2018A&A...617L...1C}, corresponding to 0.05 -- 0.15 mas based on the interferometric diameter of V1261 Ori. This would introduce a further offset, in principle, in a random direction, superimposed on the parallax and binary motion. When we included such offsets in our simulations, the conclusions remained largely unchanged, suggesting that this effect is rather small. To fully resolve the discrepancy, additional factors, such as modifications to the Roche-lobe potential or the stellar wind, have to be considered, as discussed in Sect. \ref{sec:ori}.

        \begin{figure}
   \centering
   \includegraphics[width=0.9\columnwidth]{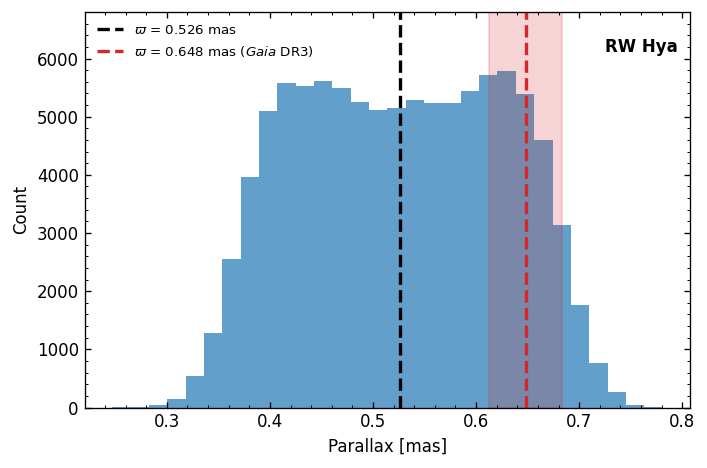}

      \caption{
      Same as Fig. \ref{fig:v1261ori_astrometry}, but for RW Hya.}

         \label{fig:rwhya_astrometry}
   \end{figure}

\subsection{Case of RW Hya}

We applied the same analysis to RW Hya, adopting a distance of 1\,900 pc ($\varpi$ = 0.526 mas), at which the giant could fill its Roche lobe (see Sect. \ref{sec:rwhya}). In this case, the orbital period is very close to one year (370.4 days; Table \ref{tab:target_stars}). Depending on the orientation and time of periastron passage, the binary motion could either amplify the parallax by a factor of $\approx$1.4 or counteract it, significantly reducing the measured value (Fig. \ref{fig:rwhya_astrometry}). Unlike V1261 Ori, the \textit{Gaia} DR3 parallax for RW Hya lies comfortably within the range of values expected when binary motion is not accounted for. Thus, in this case, an incorrect distance due to binarity can plausibly explain the observed inconsistency between the Roche-lobe filling factor and the ellipsoidal variability.

\subsection{General symbiotic population}
To evaluate the impact of binarity on astrometric solutions in the case of symbiotic stars in general, we repeated the analysis of \citet{2020MNRAS.495..321P}, focusing not on a general population of binaries but on a subset with parameters typical of symbiotic systems. We simulated astrometric observations using the actual \textit{Gaia} DR3 scanning law for 500\,000 binaries. Each binary was placed at a random position in the sky, with a randomly oriented orbit and time of periastron passage. Their magnitudes were drawn from a normal distribution with a mean of $G$ = 13 mag and a standard deviation of 1.8 mag, closely matching the \textit{Gaia G} magnitude distribution of known galactic symbiotic stars from the New Online Database of Symbiotic Variables \citep{2019AN....340..598M,2019RNAAS...3...28M}. Proper motions were sampled between -10 and 4 mas/yr in each coordinate, consistent with the typical motions of known symbiotic stars. We tested input parallaxes ranging from 0.1 to 2.2 mas. Orbital periods from 200 to 5000 days were considered, and mass ratios were drawn from a normal distribution with a mean of 2.5 and a standard deviation of 0.4, reflecting the typical values for symbiotic stars (with the exception of a small subset having q < 1). Luminosity ratios were taken from a uniform distribution ranging from 0.005 to 0.15. If the actual luminosity ratio is higher (i.e., if radiation from the hot component and/or nebula contributes more significantly in the \textit{Gaia} G band than the tested range), the results should be interpreted as upper limits.

\begin{figure*}
 
   \centering
   \includegraphics[width=0.86\columnwidth]{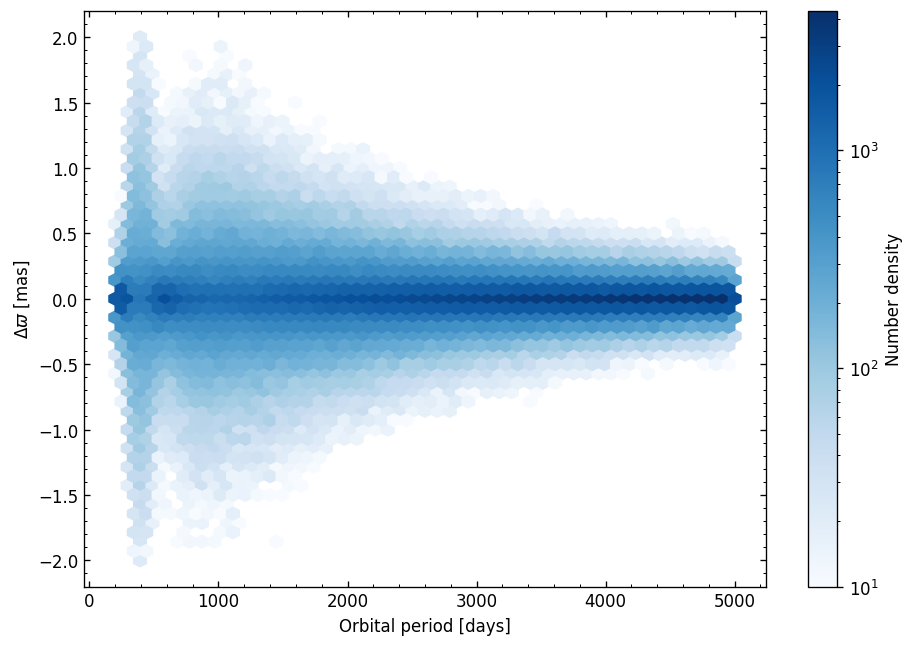}
   \includegraphics[width=0.86\columnwidth]{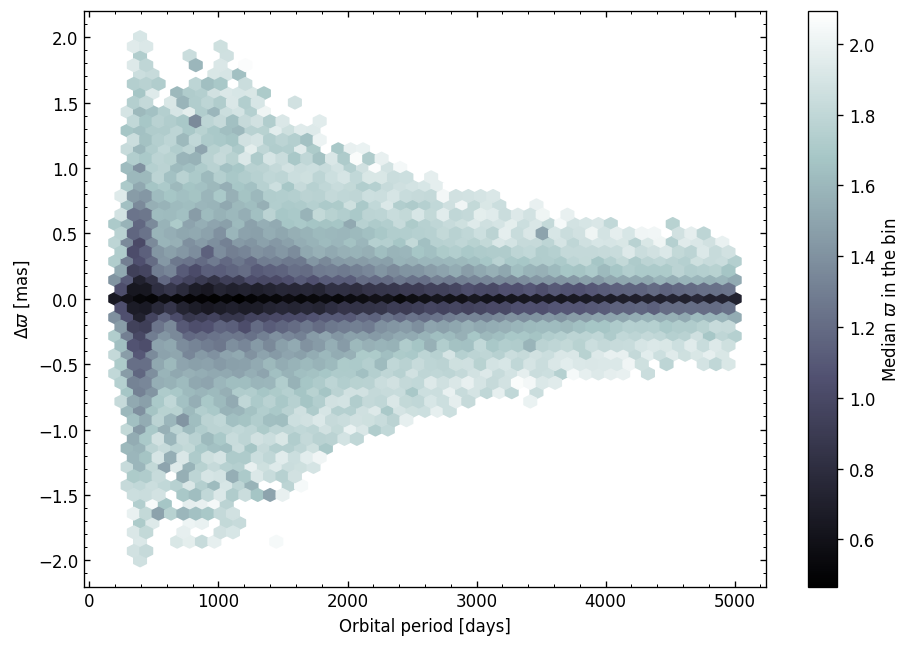}
   \includegraphics[width=0.86\columnwidth]{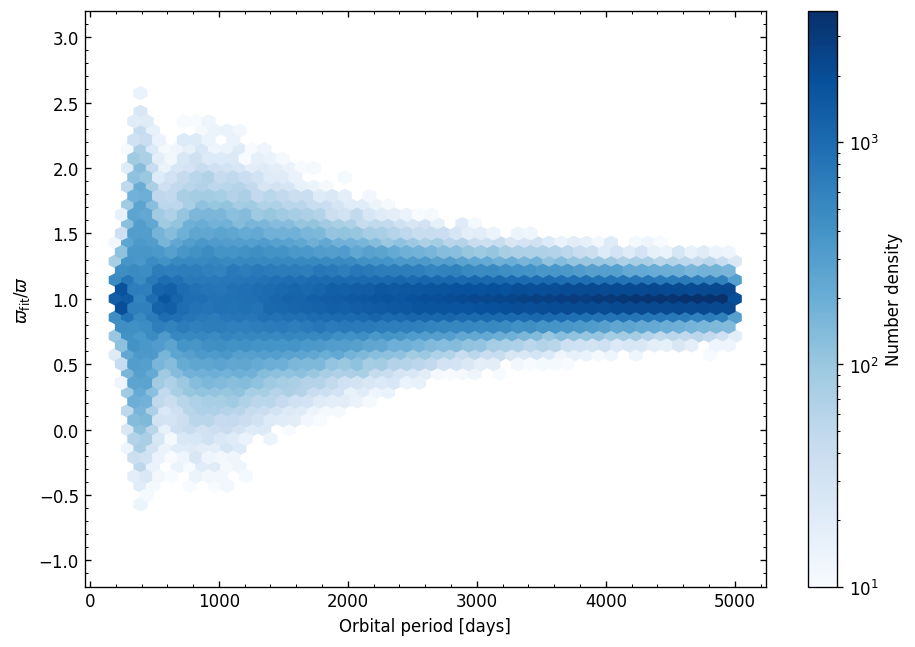}
   \includegraphics[width=0.86\columnwidth]{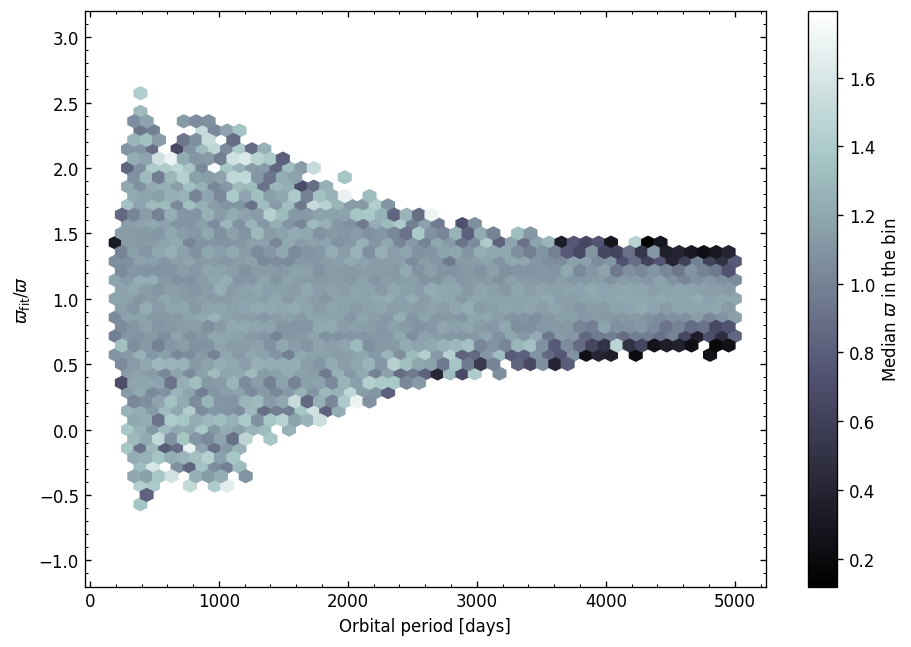}
  
      \caption{Parallax deviation as a function of the orbital period. Upper panels: Absolute differences between the input and fitted parallaxes. Lower panels: Ratio between these values. Panels on the left show the number density, while the panels on the right are color-coded by the median parallax in the particular bin. }
         \label{fig:gaia_binarity1}
   \end{figure*}

\begin{figure*}
 
   \centering
   \includegraphics[width=0.86\columnwidth]{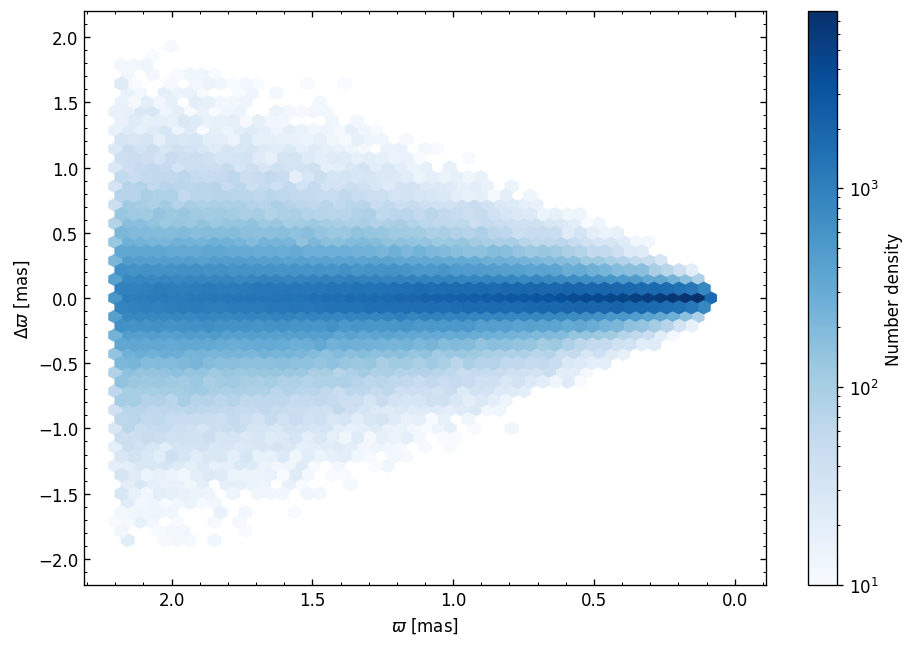}
   \includegraphics[width=0.86\columnwidth]{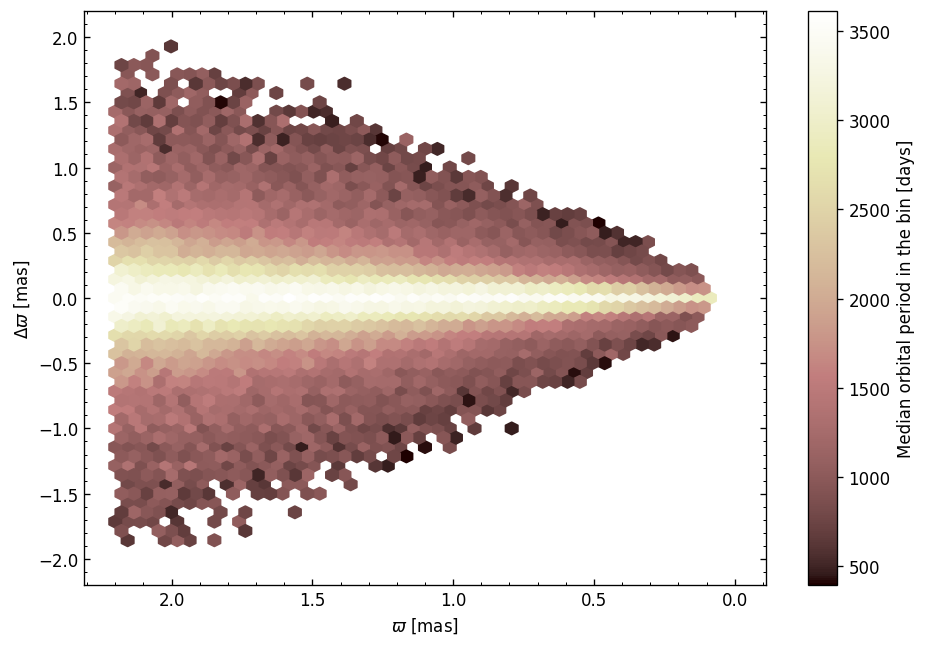}
   \includegraphics[width=0.86\columnwidth]{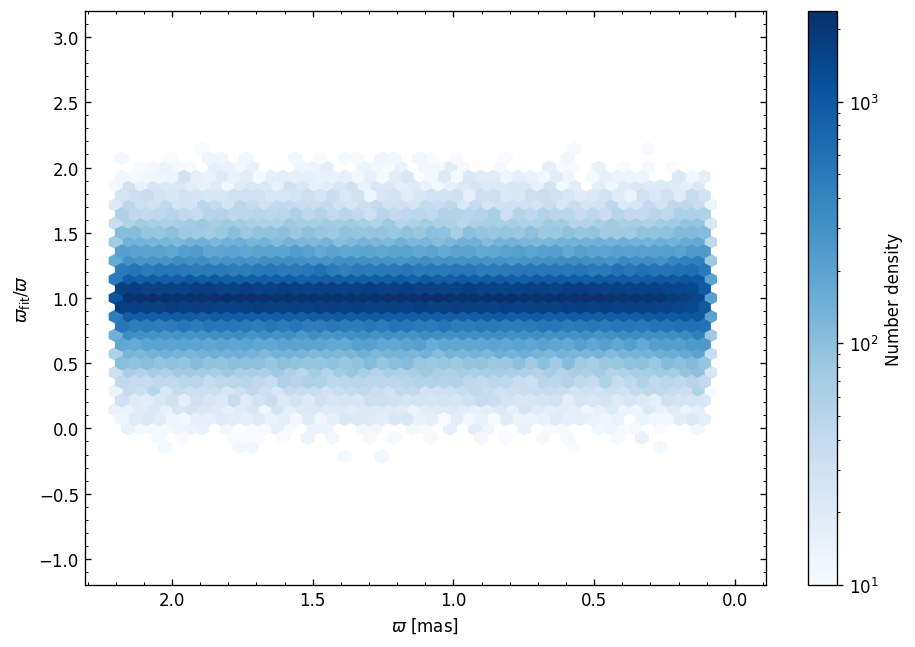}
   \includegraphics[width=0.86\columnwidth]{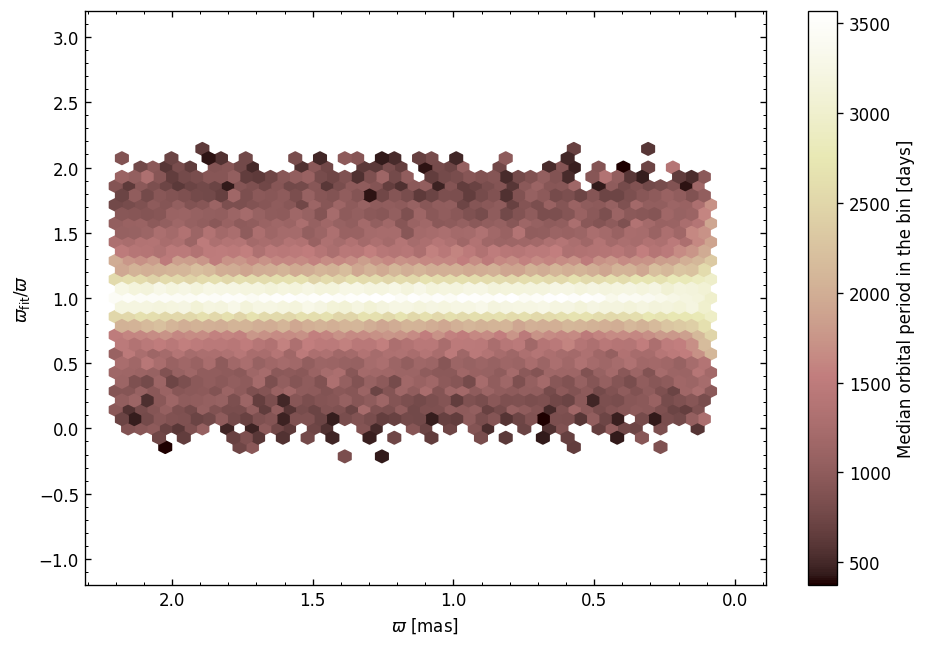}
  
      \caption{Parallax deviation as a function of the parallax (i.e., distance) for absolute (upper) and relative deviations (lower). The left panels show the number density, and the right panels are color-coded with the median orbital period in the particular bin. }
         \label{fig:gaia_binarity2}
   \end{figure*}

The impact of binarity on the inferred parallax for this sample of binaries is illustrated in Fig. \ref{fig:gaia_binarity1} (as a function of the orbital period) and Fig. \ref{fig:gaia_binarity2} (as a function of parallax/distance). As expected, the parallax is most affected when the binary period is close to one year, with the largest absolute differences reaching up to 1.5 -- 2 mas. The top panels of Fig. \ref{fig:gaia_binarity1} show that the absolute difference is, on average, higher for closer objects (indicated by the color-coding, which represents the median parallax in each bin) compared to more distant binaries. This is because the apparent size of the binary orbit on the sky decreases with increasing distance.

To provide additional context, the relative differences ($\varpi_{\rm fit}/\varpi$) are also plotted (bottom panels of Fig. \ref{fig:gaia_binarity1}), demonstrating that parallaxes can be influenced by up to a factor of $\approx$2, regardless of distance. The effect is less pronounced for binaries with longer orbital periods. Figure \ref{fig:gaia_binarity2} presents the same data but in a complementary manner, with plots color-coded by orbital period, offering a different perspective on the trends.

 \begin{figure*}
 \section{Additional figures}

   \centering
   \includegraphics[width=1\columnwidth]{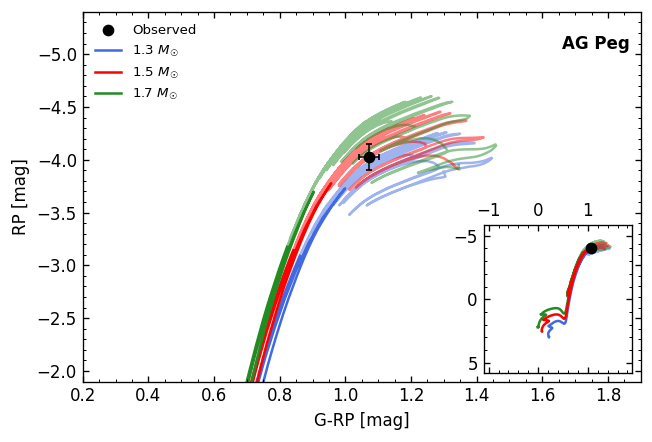}
   \includegraphics[width=0.99\columnwidth]{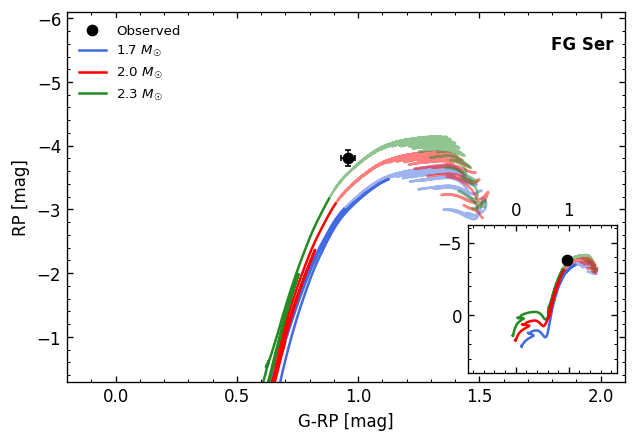}
   \includegraphics[width=1\columnwidth]{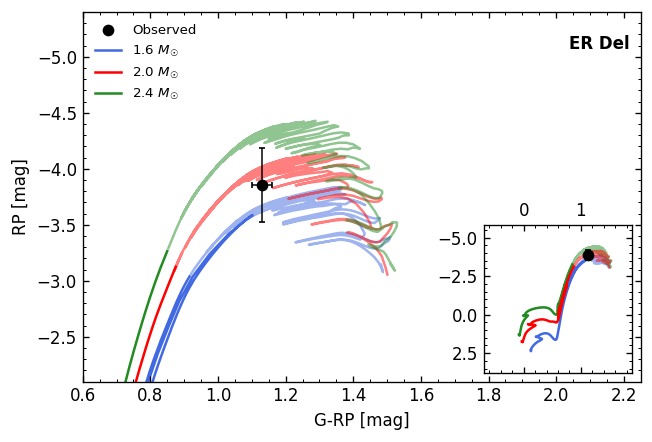}
   \includegraphics[width=1\columnwidth]{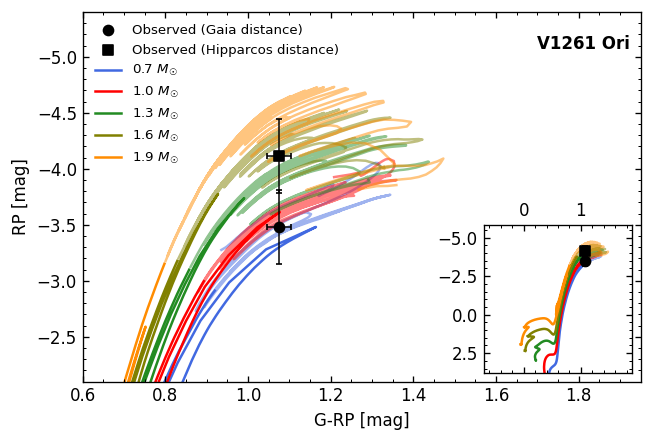}
   \includegraphics[width=1\columnwidth]{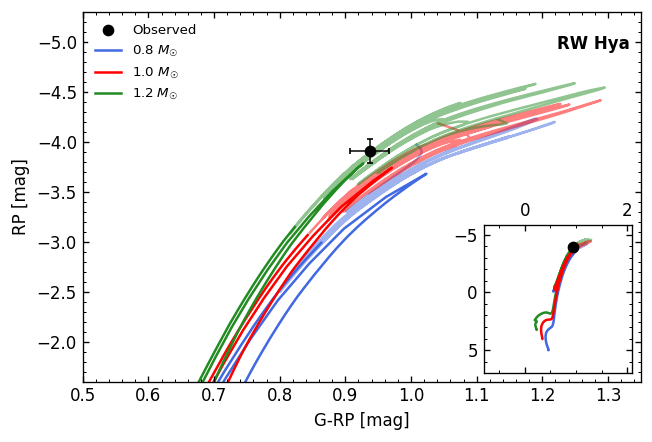}
   \includegraphics[width=1\columnwidth]{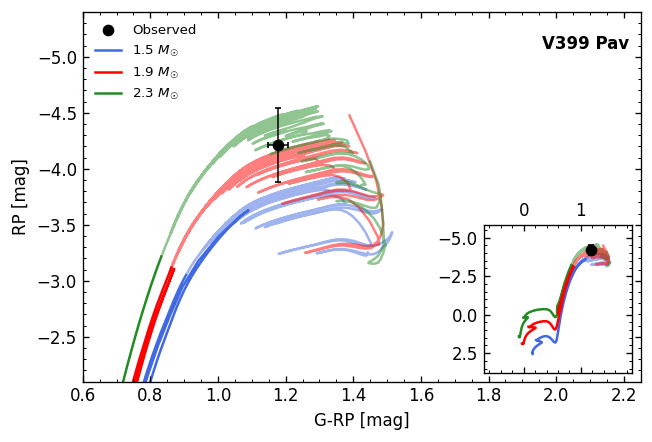}

      \caption{Position of target stars in the \textit{Gaia} color-magnitude diagram. The absolute \textit{Gaia} RP magnitudes are calculated using parallax-based distances from Table \ref{tab:gaia}. Magnitudes are corrected for interstellar extinction (details provided in the text), but no corrections are made for the contribution of additional radiation sources within the symbiotic system (e.g., the symbiotic nebula, which may cause the stars to appear bluer and brighter). The evolutionary tracks correspond to the same initial masses as shown in Fig.~\ref{fig:hr}.
              }
         \label{fig:cmd}
   \end{figure*}

 \begin{figure*}
 
   \centering
   \includegraphics[width=0.98\columnwidth]{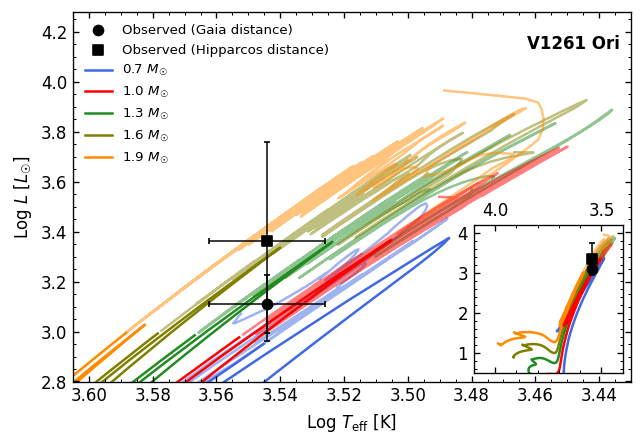}
   \includegraphics[width=1\columnwidth]{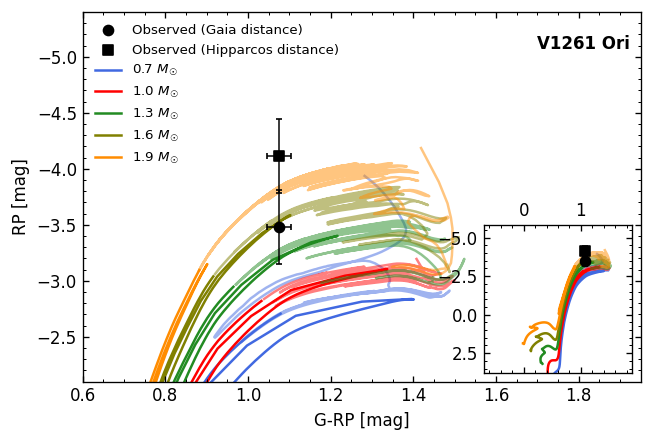}
  
      \caption{Position of V1261 Ori in the H-R diagram (\textit{left}) and \textit{Gaia} color-magnitude diagram (\textit{right}). Panels are same as Figs.~\ref{fig:hr} and \ref{fig:cmd}, but adopting the metallicity of -0.18 dex at the upper limit of the uncertainties of \citet{2016A&A...585A..64S} for V1261 Ori.
              }
         \label{fig:v1261ori_alternative}
   \end{figure*}

\end{appendix}

\end{document}